\documentclass[11pt]{article}

\usepackage[top=2cm,bottom=2cm,left=2cm,right=2cm,a4paper]{geometry}

\usepackage{amsfonts, amsmath, amsthm, graphicx}

\usepackage[T1]{fontenc}

\usepackage{color}

\usepackage{hyperref}

\usepackage{authblk}

\definecolor{reddish}{rgb}{0.75,0.2,0.2}

\definecolor{linkc}{rgb}{0,0,0.3}
\definecolor{rule}{rgb}{0.7,0.7,0.7}

\hypersetup{colorlinks=true, linkcolor=black, urlcolor=linkc, citecolor=linkc, bookmarksopen}

\newcommand{\dsty}{\displaystyle}

\newcommand{\bibm}{\textsc{bib}}
\newcommand{\hl}{\textsc{hl}}
\newcommand{\cdt}{\textsc{cdt}}
\newcommand{\regge}{\textsc{regge}}

\newcommand{\extd}{\textrm{d}}

\newcommand{\be}{\begin{equation}}
\newcommand{\ee}{\end{equation}}
\newcommand{\f}{\frac}

\begin{document}


\title{\bf Capturing the phase diagram of (2+1)-dimensional CDT using a balls-in-boxes model}

\author[1]{Dario Benedetti}
\author[2]{James P.~Ryan}

\affil[1]{\normalsize\it Laboratoire de Physique Th\'eorique (UMR8627), CNRS, Univ.Paris-Sud, \authorcr
\it Universit\'e Paris-Saclay, 91405 Orsay, France \authorcr
email: dario.benedetti@th.u-psud.fr  \authorcr \hfill }

\affil[2]{\normalsize\it Institute for Mathematics, Astrophysics and Particle Physcs, \authorcr 
    \it Radboud University, Nijmegen, The Netherlands \authorcr
email: j.ryan@science.ru.nl}

\date{}

\maketitle

\begin{abstract}
\noindent We study the phase diagram of a one-dimensional \emph{balls-in-boxes} (BIB) model that has been proposed as an effective model for the spatial-volume dynamics of (2+1)-dimensional \emph{causal dynamical triangulations} (CDT). The latter is a statistical model of random geometries and a candidate for a nonperturbative formulation of quantum gravity, and it is known to have an interesting phase diagram, in particular including a phase of extended geometry with classical properties. Our results corroborate a previous analysis suggesting that a particular type of potential is needed in the BIB model in order to reproduce the droplet condensation typical of the extended phase of CDT. Since such a potential can be obtained by a minisuperspace reduction of a (2+1)-dimensional gravity theory of the Ho\v{r}ava-Lifshitz type, our result strengthens the link between CDT and Ho\v{r}ava-Lifshitz gravity.

\end{abstract}


\section{Introduction}

The emergence of spacetime and its classical dynamics from some underlying fundamental theory is the dream of most approaches to quantum gravity.
One natural attempt in such a direction is to try to build spacetime from some fundamental building blocks, following some set of ``microscopic'' rules (i.e. interactions between the building blocks), and look for the emergence of classical geometries with classical dynamics at ``macroscopic'' scales (i.e. at scales much larger than the typical scale of a building block, in a system containing a large number of them). 
Several approaches to quantum gravity fall rather directly into this category: dynamical triangulations \cite{Ambjorn:book}, causal sets \cite{Henson:2006kf}, group field theory \cite{Oriti:2011jm}, tensor models \cite{Gurau:2016cjo}, and causal dynamical triangulations \cite{Ambjorn:2012jv}. 
Among all of them, the latter stands out as the only one so far having succeeded in passing the first test, that is, producing an emergent geometry with at least some classical features (such as the correct Hausdorff dimension), and with an effective dynamics (describing the spatial-volume evolution in the ground state and the fluctuations around it) governed by a classical action, which in 3+1 dimensions is compatible with general relativity (apart from a sign \cite{Ambjorn:2004pw}).

On the other hand, there are reasons to believe that without any special fine tuning the full dynamics of CDT would not in general reproduce that of general relativity. In fact, one of the main ingredients of CDT, the one that distinguishes it from the old dynamical triangulations models and which is unique among all the other approaches mentioned above, is the presence of a preferred foliation.
This fact, together with a number of recent results \cite{Horava:2009if,Ambjorn:2010hu,Benedetti:2009ge,Anderson:2011bj,Budd:2011zm,Ambjorn:2013joa,Benedetti:2014dra}, has led to the suggestion that the effective field theory underlying CDT is not general relativity, but Ho\v{r}ava-Lifshitz gravity  \cite{Horava:2008ih,Horava:2009uw}.
Since the latter faces a number of challenges in recovering general relativity at low energy \cite{Charmousis:2009tc,Blas:2010hb,Mukohyama:2010xz}, if the link between the two were to be confirmed, this would raise some concerns about CDT as a viable theory of quantum gravity.
Or at least it would tell us that a very special fine tuning is needed when taking the continuum limit: we are free to view general relativity as a special case of HL gravity, corresponding to a very precise choice of its couplings, such that full diffeomorphism invariance is recovered; in the same way, it might be possible to fine tune CDT to restore in the continuum limit the invariance broken by the foliation.
In any case, regardless of the phenomenological viability of HL gravity or of the fine tuning of the continuum limit, if we could show in some detail that CDT admits a continuum limit described by HL gravity, this would already be a spectacular result for a theory of random geometry of this type, the only known precedent being in two dimensions the link between dynamical triangulations and Liouville gravity \cite{David:2014aha}.

Unfortunately, the (3+1)-dimensional case is very difficult to study, from both sides of the story: for that dimension, CDT has a complex phase diagram that so far we can only approach by Monte Carlo simulations, while HL gravity is a complicated theory with many couplings.
One natural option is then to consider lower dimensional models.
Things get extremely simple in 1+1 dimensions, where in fact CDT (without matter) is exactly solvable \cite{Ambjorn:1998xu}, and HL gravity becomes a one-dimensional model for the length of the spatial slices \cite{Ambjorn:2013joa}.
In this case it has been shown that indeed the continuum limit of CDT is HL gravity \cite{Ambjorn:2013joa}.
However, the (1+1)-dimensional case being rather trivial, it is not obvious that the relation between the two would generalize to higher dimensions.
An intermediate case is that of 2+1 dimensions, where both CDT and HL gravity are non-trivial, but definitely simpler than in one dimension higher.
CDT in 2+1 dimensions is so far not solvable, and its properties have also been studied mostly by numerical simulations \cite{Ambjorn:2000dja,Ambjorn:2002nu,Benedetti:2009ge,Anderson:2011bj,Cooperman:2013mma,Budd:2011zm,Budd:2013waa,Jordan:2013iaa}, but one can in principle reach larger volumes than in higher dimensions, and furthermore some analytical tools are available \cite{Ambjorn:2001br,Benedetti:2007pp} (and others might be developed more likely than in higher dimensions).
Similarly, HL gravity in 2+1 dimensions has so far not been solved, but it is a theory of only one degree of freedom, for which one-loop computations are possible \cite{Benedetti:2013pya,Barvinsky:2015kil}.

Here we want to strengthen the links between CDT and HL gravity by following directly on \cite{Benedetti:2014dra}, where the case for an effective HL-type description of (2+1)-dimensional CDT was made.
The observable discussed in that work is the time-dependent volume (or area) of the spatial slices $V_2(t)$, which one can easily compare to classical solutions of minisuperspace reductions of the target continuum theory. 
Due to the foliation, the spatial volume is a natural observable in CDT and it is the main one characterizing the different phases: in particular, the extended geometry phase corresponds to a volume profile that averages to a droplet configuration, i.e. it is characterized by one maximum, at say $t=0$, around which is concentrated the bulk of the total volume and by a stalk of minimal spatial volume (see figure \ref{fig:bib-cdt}).
We refer to this configuration also as a condensation, both because the total volume condenses in a subset of the total time interval (which is kept fixed in the simulations) and in opposition to other phases, where the volume is more evenly distributed (averaging for example to a constant).
The idea put forward in \cite{Benedetti:2014dra} is based on a very simple observation: when applying a minisuperspace reduction from the full metric $g_{\mu\nu}(\vec{x},t)$ to the scale factor $\phi(t)$ of the spatial slices, the general relativity action in 2+1 dimensions reduces to a purely kinetic term, with no potential for $\phi(t)$; on the contrary, HL gravity leads in general to a non-trivial potential.
As argued in \cite{Benedetti:2014dra}, only in the presence of a potential can droplet condensation occur. It was argued that the potential arising in HL gravity for the case of spherical spatial slices leads precisely to a dominance of droplet configurations in the path integral (for some range of parameters) and that the shape of such droplets match remarkably well the one of CDT.

However, the analysis in  \cite{Benedetti:2014dra} was of a semiclassical nature and it required a number of assumptions. 
Furthermore, the proposed model contains a stabilizing constraint that leads to a boundary in configuration space and, as a consequence, to the non-standard situation in which a configuration that is not a solution of the equations of motion dominating the path integral.  
Most importantly, while such analysis showed that a droplet configuration would dominate over a constant one for certain values of parameters, no strong evidence could be offered to exclude the presence of other configurations being even more dominant. 
In fact, the configuration discussed  in  \cite{Benedetti:2014dra} is probably just an approximation to the real mean configuration, which in particular we would expect to have a smoother behavior at the junction between the bulk of the droplet and the stalk.
All that will be reviewed and clarified in section \ref{sec:model}.

The weaknesses of the semicalssical analysis are a strong motivation for probing the non-perturbative regime of the model by different means.
This is precisely the scope of the work that we present in section \ref{sec:sim}, where we study a BIB model with the methods developed in \cite{Bogacz:2012sa}.
In particular, we have studied the BIB model via Monte Carlo simulations and clearly identified at least five different phases, among which is present a droplet phase with the properties predicted in  \cite{Benedetti:2014dra}.

A brief summary and outlook is given in section \ref{sec:concl}.

\section{Origin and analysis of the model}
\label{sec:model}

\subsection{BIB\ model}
In the context of dynamical triangulations,\footnote{To distinguish this from their use in modeling zero-range processes \cite{Evans-review}, where the weight factors are generally of the simpler form: $g(m_j)$.} a balls-in-boxes model is defined via a canonical partition function of the form:
\begin{equation}
  \label{eq:bib-pf}
  \begin{split}
  Z_{\bibm}(T,M) &= \sum_{m_0 = m_{min}}^{M}\cdots\sum_{m_{T-1} = m_{min}}^{M} \delta_{M,\sum_j m_j} \prod_{j = 0}^{T-1} g(m_j, m_{j+1})\\
  & = \sum_{\{m_j\}} e^{-S[\{m_j\}]}\delta_{M,\sum_j m_j}\,.
  \end{split}
\end{equation}
Such a system has support on a one-dimensional circular lattice with $T$ sites (circular in the sense that the sites are indexed modulo $T$). To the $j$-th site, one associates an integer $m_j \ge m_{min} > 0$, such that their sum is subject to the constraint:\footnote{As one might imagine, the lattice sites and integers correspond to the aforementioned \emph{boxes} and \emph{balls}, respectively. Meanwhile, equation \eqref{eq:bib-constraint} constrains the total number of balls.}
\begin{equation}
  \label{eq:bib-constraint}
  \sum_{j = 0}^{T-1} m_j = M\,.
\end{equation}
The weight factors $g(m_j,m_{j+1})$ capture a nearest-neighbour interaction and if it was not for the delta function imposing the constraint \eqref{eq:bib-constraint} the partition function \eqref{eq:bib-pf} would be the standard discretized path integral for a one-dimensional model.
In the case proposed in \cite{Benedetti:2014dra}, the precise form of the weight factors is:
\begin{equation}
  \label{eq:bib-2plus1}
  g(m_j,m_{j+1}) = \exp\left\{-\frac{2(m_{j+1} - m_j)^2}{m_{j+1} + m_j}b_1 + \frac{2}{m_{j+1} + m_j}b_2\right\}\,,
\end{equation}
where $(b_1,\,b_2)$ are phase space parameters, and the factors 2 remind us that we have chosen to write in the denominators the arithmetic mean of $m_j$ and $m_{j+1}$ (other choices, even non-symmetric, are possible, but are expected to be irrelevant in the continuum limit \cite{Bogacz:2012sa}).  This is the model under investigation hereafter.

\begin{figure}[h]
    \begin{center}
        \includegraphics[scale = 0.45]{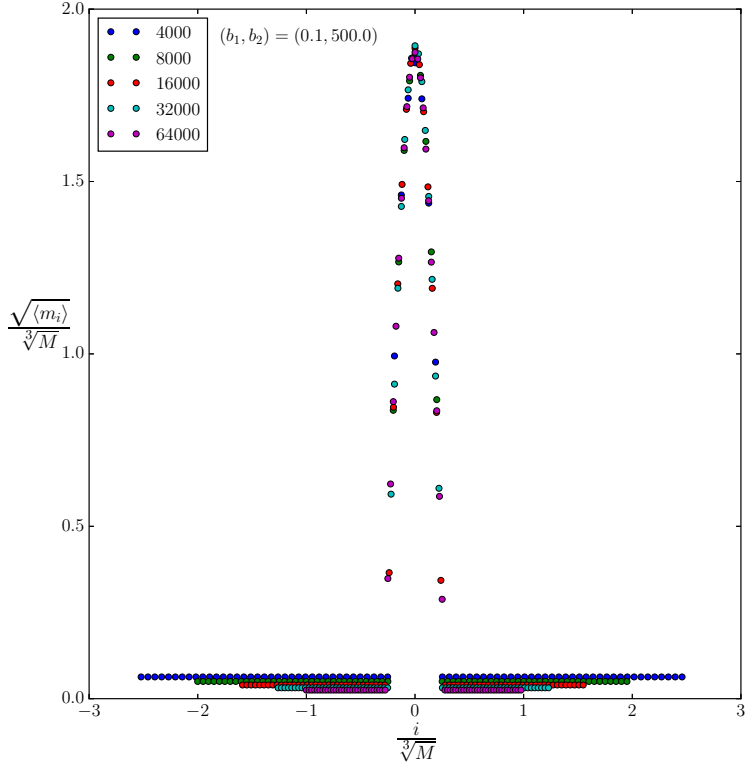}
        \caption{\label{fig:scaling} \small{Scaling of averaged droplet with respect to volume $M$ at fixed $T = 80$. We notice that rescaling both $\sqrt{\langle m_i\rangle}$ and $i$ with $\sqrt[3]{M}$, we obtain a droplet whose shape is essentially independent of $M$.}}
    \end{center}
\end{figure}

At large $T$ and $M$ we can use a continuum approximation, introducing a lattice spacing $a$, and defining a continuous time variable  $t= a j$, with period $\tau=a T$. Furthermore, we will assume a three-dimensional interpretation of the model, which is justified a posteriori by the scaling observed in the droplet phase (see figure \ref{fig:scaling}), therefore defining a continuous volume variable  $V_3=v_3 a^3 M$, as well as interpreting the number of balls in a given box $j$ as the two-dimensional volume of a slice at time $t$: $V_2= v_2 a^2 m$.
Here, $v_2$ and $v_3$ are two arbitrary constants.

In the limit $a\to 0$, with $\tau$ and $V_3$ fixed, the action $S$ in \eqref{eq:bib-pf} becomes:
\be \label{eq:bib-contS}
S[V_2] = \frac{1}{a} \int_0^\tau dt \left\{ \frac{b_1}{v_2} \frac{\dot{V_2}^2}{V_2} - a^2 b_2 v_2 \frac{1}{V_2} \right\} \, ,
\ee
while the constraint \eqref{eq:bib-constraint} becomes:
\be
\int_0^\tau dt V_2 = \frac{v_2}{v_3} V_3 \, .
\ee
It is then natural to take $v_2=v_3\equiv v$.

Note that the presence of the $1/a$ factor in front of the action suggests that a good way to minimize  the action is to identify the behavior of the model in the continuum limit (but at finite $a$, i.e. at finite volume, one should take also $b_1$ and $b_2$ not too small for the same analysis to apply).

\subsection{From gravity to balls-in-boxes}
\label{ssec:hl-to-bib}

We can reverse the logic of the previous section and, starting from a gravity action in the continuum, we can reduce it to a BIB model in the discrete.

\textbf{Starting point:} One sets off with the most generic, projectable, $z = 2$ action for Ho\v rava-Lifshitz gravity in $(2+1)$ dimensions:
\begin{equation}
  \label{eq:hl-action}
  S_{(2+1)-\hl} = \frac{1}{16\pi G} \int \extd t\,\extd^2x\; N\sqrt{g}
                  \left\{
                      \sigma(\lambda K^2 -  K_{ij}K^{ij})
                      - 2\Lambda
                      + bR
                      - \gamma R^2
                  \right\}\,.
\end{equation}
We shall return in Section \ref{ssec:hl-to-cdt} to detail why one might chose this starting point but, for the moment, let us just present a digest of the derivation leading to the BIB\ model \eqref{eq:bib-2plus1}.

HL\ gravity describes a class of metric theories supporting a preferred foliation. Thus, the more familiar space-time diffeomorphism symmetry is broken down to foliation-preserving diffeomorphisms.\footnote{Such foliation-preserving diffeomorphisms are described by time-reparameterizations and spatial diffeomorphisms of the form:
  \begin{equation}
    \label{eq:hl-diffeos}
    t \rightarrow t + \xi^0(t)\qquad\textrm{and}\qquad x^i \rightarrow x^i + \xi^i(t,x)\,,
  \end{equation}
  where $(t,x^i)$ are co-ordinates in an atlas of charts adapted to the foliation.
}

In \eqref{eq:hl-action}, the action is presented in terms of ADM variables: $N$ is the lapse function, $g$ is the determinant of the spatial metric, $R$ is its Ricci scalar, $K_{ij}$ is the extrinsic curvature associated to the leaves of the foliation, while $K$ is its trace.

The action contains the familiar parameter pair $(G, \Lambda)$, corresponding to Newton's constant and the cosmological constant, respectively. The parameter $\sigma = \pm1$ neatly encapsulates some metric signature information.  Meanwhile $\lambda$, $b$ and $\gamma$ characterize the deviation from full diffeomorphism invariance. Indeed, for $\lambda = b = 1$ and $\gamma = 0$, one recovers the Einstein-Hilbert action with either Euclidean ($\sigma = 1$) or Lorentzian ($\sigma = -1$) signature.\footnote{In both cases we obtain the Lagrangian $\mathcal{R}-2\Lambda$, where $\mathcal{R}$ is the spacetime Ricci tensor. Notice that the action was chosen to have in this case an overall minus sign with respect to standard GR, in order to match the sign observed in CDT simulations \cite{Ambjorn:2002nu,Ambjorn:2004pw}. In fact CDT has long been known to lead to a positive sign for the kinetic term for the conformal mode, thus avoiding the famous conformal sickness of GR.}

The theory is said to be \textit{projectable} if one imposes at the outset that the lapse function is spatially constant: $N = N(t)$.

The $z$-\textit{exponent} refers to half the maximal order of spatial derivatives arising in the inverse propagator of the free theory. Thus for $z = 2$, it may contain at most quartic terms, permitting the inclusion of the $R^2$ term.

\textbf{Space-time topology and boundary conditions:} We consider spacetimes with topology $S^{2}\times S^1$, that is, spherical spatial slices and a compactified time. To implement this, we impose periodic boundary conditions in time, with period $\tau$.

\textbf{Mini-superspace reduction:} We perform a mini-superspace reduction to constant lapse, vanishing shift vector (hidden so far in the extrinsic curvature) and spatial metric $g_{ij} = \phi^2(t) \hat{g}_{ij}$, where $\hat{g}_{ij}$ is the standard metric on the unit sphere.  The action becomes:\footnote{The redefinition of parameters goes as follows:
  \begin{equation}
    \label{eq:hl-mini-parameters}
    \kappa^2 = \sigma \frac{NG}{2\lambda - 1},\qquad
    \omega^2 = \sigma\frac{N^2\Lambda}{2\lambda -1}\,,\qquad
    b' = \sigma\frac{N^2b}{2\lambda -1}\,,\qquad
    \xi = \sigma\frac{2N^2\gamma}{2\lambda -1}\,.
  \end{equation}
}
\begin{equation}
  \label{eq:hl-mini-action}
  S_{(2+1)-\hl-\textrm{mini}} = \frac{1}{2\kappa^2}\int_{-\frac{\tau}{2}}^{\frac{\tau}{2}} \extd t \left\{\dot{\phi}^2 - \omega^2\phi^2 + b' - \frac{\xi}{\phi^2}\right\}\,.
\end{equation}
The remaining field $\phi(t)$ is a time-dependent scale factor determining the area of the spatial slice at time $t$: $V_2(t) = 4\pi\phi^2(t)$.
The constant $b'$ term is clearly irrelevant for the problem of minimization of the action, and this can be traced back to the topological nature of the $R$ term in \eqref{eq:hl-action} for the projectable case.

For concreteness of the analysis, we take $\sigma(2\lambda - 1) > 0$.  As a result, the kinetic term is positive definite.
However, in order to satisfy the periodic boundary conditions and to get real oscillating solutions, it turns out that one needs to take $\omega^2 > 0$ (that is, a positive bare cosmological constant) and $\xi > 0$ (that is, an  $R^2$ term with the wrong sign), thus leading to a potential which is unbounded from below. Both sources of unboundedness can be cured by constraining the configuration space:

\begin{enumerate}
  \item One can avoid  the potential unboundedness stemming from the $1/\phi^2$ term by imposing a minimal spatial volume constraint at the outset, such as:
\begin{equation}
  \label{eq:minimal-constraint}
  \phi(t) >\epsilon\,,\quad \forall t\,.
\end{equation}
This can be thought as a regularization of the theory, and as discussed in  \cite{Benedetti:2014dra}, there are possible scaling limits to safely let $\epsilon\to 0$.

\item The unbounded $\phi^2$ term is tamed by the working in the canonical ensemble, i.e. at fixed total volume. To that end, we add a term $\frac{\omega^2}{8\pi N \kappa^2}V_3$, where $V_3$ is a constant, to \eqref{eq:hl-mini-action} and treat $\omega^2$ as a Lagrange multiplier.  Such an addition leaves the existing Euler-Lagrange equations unchanged, while variation with respect to the Lagrange multiplier imposes the constraint:
\begin{equation}
  \label{eq:hl-volume-constraint}
  \mathcal{V} \equiv 4\pi N\int_{-\frac\tau2}^{\frac\tau2} \extd t \, \phi^2(t) - V_3 = 0\,,
\end{equation}
that is, it fixes the 3-volume.

Note that the compatibility of \eqref{eq:hl-volume-constraint} with \eqref{eq:minimal-constraint} requires $\epsilon<\sqrt{ V_3/(4\pi N\tau) }$.

\end{enumerate}

Note also that the constant lapse $N$ has been neatly hidden away in  \eqref{eq:hl-mini-action}. This is rather innocuous in a classical setting. But in principle, it is a degree of freedom that should be integrated over in a quantum regime leading to the imposition of a Hamiltonian constraint. Following \cite{Benedetti:2014dra}, we shall not consider this scenario, rather setting $N=1$ from here on. As we shall see in the Section \ref{ssec:hl-to-cdt}, that is appropriate when wishing to set it in comparison with CDT.

In order to complete the link with \eqref{eq:bib-contS}, one should simply change variables to:
\be
V_2 = 4\pi \phi^2\,,
\ee
which comes from having assumed the slices to be 2-spheres of radius $\phi$.
Leaving aside the constant term and the harmonic term, which is part of the volume constraint, one can rewrite \eqref{eq:hl-mini-action} as:
\be \label{eq:action}
 \tilde S_{(2+1)-\hl-\textrm{mini}} \equiv \frac{1}{2\kappa^2}\int_{-\frac{\tau}{2}}^{\frac{\tau}{2}} \extd t \left\{\dot{\phi}^2  - \frac{\xi}{\phi^2}\right\} =  \frac{1}{2\kappa^2}\int_{-\frac{\tau}{2}}^{\frac{\tau}{2}} \extd t \left\{\frac{1}{16\pi}\frac{\dot{V_2}^2}{V_2}  - \frac{4\pi\xi}{V_2}\right\}\, \, ,
\ee
which upon comparison with \eqref{eq:bib-contS}  leads to the identifications:
\be
\frac{1}{32\pi\kappa^2} = \frac{b_1}{a v} \,, \;\;\;   \frac{2\pi\xi}{\kappa^2}= a b_2 v \, ,
\ee
or:
\be
\frac{b_2}{b_1} = \frac{64 \pi^2 \xi}{v^2 a^2} \, .
\ee
One can eliminate all references to the cutoff scale $a$ by looking only at dimensionless ratios of the parameters of the model. For example, using $v a^2 m_{min} = 4\pi \epsilon^2$, one can write:
\be
\frac{v\, b_2}{b_1 m_{min}} = \frac{16 \pi \xi}{\epsilon^2} \, .
\ee
In this way we will be able to translate our predictions from the continuum to the discrete.

Notice that general relativity corresponds to the case $b_2=0$.

\subsection{Analysis of the model}
\label{ssec:analysis}

The equations of motion derived from \eqref{eq:hl-mini-action} are of Ermakov-Pinney type, i.e.:
  \begin{equation}
    \label{eq:hl-pe-equations}
    \ddot\phi + \omega^2\phi - \frac{\xi}{\phi^3} = 0\,,
  \end{equation}
and their solution is:
\begin{equation}
  \label{eq:hl-pe-solutions}
  \phi_0(t) = \frac{1}{\omega A} \sqrt{(\omega^2A^4 - \xi)\cos^2(\omega t + \psi) + \xi}\,,
\end{equation}
where $A$ and $\psi$ are integration constants. Shifting the maximum of the curve to $t = 0$ fixes $\psi = 0$. Thereafter, this maximal value $\phi_0(0) = A$ is fixed by initial conditions.  Meanwhile, the constraint $\mathcal{V}$ suffices to determine $\omega$ in terms of $\{V_3, A, \xi\}$. However, notice that these solutions oscillate with period $\frac{\pi}{\omega}$, and therefore in order to satisfy periodic boundary conditions at $t=\pm \tau/2$, $\omega$ must also satisfy  $\frac{\pi}{\omega} = \frac{\tau}{n}$ for some positive integer $n$. As a consequence, the space of solutions forms a discrete set.

These solutions never reach zero for $\xi > 0$. This has ramifications for the space-time topology.  In particular, it indicates the non-occurence of potential conical singularities, rather there is a minimal throat, at which the space-time bounces.

Besides the oscillating solutions, there is also a constant solution:
\begin{equation}
  \label{eq:hl-const-solution}
  \bar\phi_0(t) 
  =\sqrt{ \frac{V_3}{4\pi \tau} }\,,
\end{equation}
which is a special case of equation \eqref{eq:hl-pe-solutions}  with $A = \xi^{1/4}/\sqrt{\omega}$, and $\omega=4\pi\tau\sqrt{\xi}/V_3$ fixed by the volume constraint. This solution has special significance, since in comparison to the other local extrema, it has least action:\footnote{The on-shell action evaluates to:
  \begin{equation}
    \label{eq:hl-on-shell-explicit}
    \begin{array}{rcl}
      \tilde S_{(2+1)-\hl-\textrm{mini}}[\phi_0] &=&  \dfrac{n\pi}{8\kappa^2}\left(\dfrac{nV_3}{N\tau^2} - 8\sqrt{\xi}\right)\,,\\[0.5cm]
 \tilde S_{(2+1)-\hl-\textrm{mini}}[\bar\phi_0]     &=& -\dfrac{2\pi N\tau^2\xi}{\kappa^2V_3}\;.
\end{array}
\end{equation}
}
\begin{equation}
  \label{eq:hl-on-shell}
 \tilde S_{(2+1)-\hl-\textrm{mini}}[\bar\phi_0] \leq  \tilde S_{(2+1)-\hl-\textrm{mini}}[\phi_0] \, .
\end{equation}

Keeping in mind that droplet-stalk configurations are of interest to us, we notice that the solutions \eqref{eq:hl-pe-solutions} and \eqref{eq:hl-const-solution} above capture fairly well the bulk of the droplet and the stalk, respectively. Therefore, with a touch of naivet\'e, let us graft the two together, and construct a configuration of the form:
\begin{equation}
  \label{eq:hl-graft}
  \tilde\phi(t) = \left\{
    \begin{array}{lcl}
      \dfrac{1}{\omega A}\sqrt{(\omega^2A^4 - \xi)\cos^2(\omega t) + \xi} & \textrm{for} & t \in [-\frac{\pi}{2\omega},+\frac{\pi}{2\omega}]\,, \\[0.3cm]
      \dfrac{\sqrt{\xi}}{\omega A} & \textrm{for} & t\in[-\frac{\tau}{2},-\frac{\pi}{2\omega}) \cup (+\frac{\pi}{2\omega},+\frac{\tau}{2}]\,, \\
    \end{array}
      \right.
\end{equation}
where one period of a configuration of the form \eqref{eq:hl-pe-solutions} is confined to a subinterval of length $\pi/\omega < \tau$, while a constant configuration is attached in the remainder.

Such configurations are unusual for at least two reasons. First, they do not in general correspond to solutions of the Euler-Lagrange equations, except in the degenerate cases $A^2=(V_3+\sqrt{V_3^2-16\tau^4\xi})/(4\pi \tau)$, for which $\omega=\pi/\tau$, i.e. there is no stalk, and $A^2 = \xi^{1/2}/\omega=V_3/(4\pi \tau)$, for which $\tilde\phi = \bar\phi_0$, i.e. there is no droplet. Second, they are only $C^1([-\tau/2,\tau/2])$, the second derivative being discontinuous at $t=\pm \frac{\pi}{2\omega}$.
Nevertheless, we will now argue that they are relevant for the analysis of the model.

\textbf{Local extrema vs absolute minima:}
The partition function of the system, for the moment in the continuum, reads:
\begin{equation}
  \label{eq:hl-mini-pf}
  Z_{(2+1)-\hl-\textrm{mini}} = \int_{\phi(t)>\epsilon}\mathcal{D}\phi(t)\,
                                \delta(\mathcal{V})\,
                                \exp\left\{
                                -\frac{1}{2\kappa^2}
                                \int_{-\frac\tau2}^{\frac\tau2}\extd t
                                \left[\dot\phi^2 - \frac{\xi}{\phi^2}\right]
                                \right\}\,.
\end{equation}

In the limit $\kappa\to 0$ we expect the partition function (and the observables) to be dominated by those configurations that minimize the action.
Under the assumption that the space of field configurations is such that the action is at least differentiable (in the functional sense), there are two possible ways minima can arise: local extrema (i.e. solutions of the equations of motion) or minima lying on the boundary of configuration space.  In the case that local extrema provide the dominant configurations for the path integral, this would imply dominance of the constant solution, because as we have seen this has the least action among all the solutions. On the other hand, if a configuration lying on the boundary of configuration space provides a global extremum, then we can have a non-constant profile.  Such a possibility arises quite naturally in situations where an action unbounded from below is tamed by constraints.
In our case, as we have already pointed out, the unboundedness of the action \eqref{eq:action} at $\xi>0$ is tamed by the constraint \eqref{eq:minimal-constraint}, which introduces a boundary in configuration space. Therefore, global minima that are not local extrema can be expected.
Indeed, it was argued in \cite{Benedetti:2014dra} that this is precisely what gives rise to the droplet phase, with a profile of the form \eqref{eq:hl-graft} dominating the partition function.

The rationale behind  \eqref{eq:hl-graft} is based on a balance between kinetic and potential terms in the action (created by their relative sign difference). If one looks at the kinetic term alone (at positive $b_1$, i.e. $\kappa^2$), then it is obvious that the minimizing configuration is a constant one, $\dot\phi(t)=0$. The volume constraint then fixes the value of $\phi(t)$ to \eqref{eq:hl-const-solution}.
On the other hand, if one looks only at the potential term, then minimization would seem to push to a configuration with $\phi(t)=0$, leading to a singular value of the action which would clearly dominate over any other configuration (as long as $\dot\phi(t)$ stays finite).  This is of course the instability caused by the unboundedness of the action, and as already repeatedly stressed it is cured by the constraint \eqref{eq:minimal-constraint}, as a consequence of which the potential actually favors the configuration $\phi(t)=\epsilon$. In fact, the volume constraint forbids a solution with $\phi(t)=\epsilon$ everywhere, but clearly if it were just for the potential term a delta-function configuration $V_2(t)= (V_3-4\pi \epsilon^2\tau)\,  \delta(t-t_0) + 4\pi \epsilon^2$ would do the job.\footnote{Our choice of working with the variable $\phi$ rather than $V_2$ would not be the best in this case due to the need of taking the square root of the delta function.}

Therefore, for $\epsilon$ small enough, one sees a competition between the kinetic term, which favors the constant configuration, and the potential, which favors the delta-function configuration. What  \eqref{eq:hl-graft} envisages is a situation where the latter essentially wins,  with the constant part being fixed at the minimal slice volume for the optimal choice of $A$, but with the kinetic term smoothing out the delta function, to a shape determined by local minimization of the action.\footnote{Notice that for negative $b_1$ the kinetic term looses its smoothing effect and we might expect the delta function configuration to dominate. This is indeed what we will find in the next section (see the localized phase).}

\textbf{Phase transition:}
The discussion above was intended to provide a simple intuition of why a configuration such as \eqref{eq:hl-graft} might dominate the path integral, but of course one has to explicitly check whether that is the case. The presence of several parameters and the fact that \eqref{eq:hl-volume-constraint} leads to a cubic equation for $\omega$ complicate things, and in \cite{Benedetti:2014dra} only a perturbative analysis for small $\epsilon$ and $\xi$ was presented.

Indeed plugging \eqref{eq:hl-graft} into \eqref{eq:hl-volume-constraint} we find:
\be \label{V3constr}
V_3 = 4\pi  \left(  \f{ \pi (\xi+A^4\omega^2) }{2 A^2 \omega^3} +\f{\xi}{A^2\omega^2} \left(\tau-\f{\pi}{\omega}\right)\right)\, ,
\ee
which, multiplied by $\omega^3$ (non-zero otherwise we would have the constant solution again), gives us a cubic equation for $\omega$. It is convenient to rewrite the latter in terms of $\epsilon$ rather than $A$, taking $A=\sqrt{\xi}/(\epsilon\omega)$, because we expect the constant part to reach its minimal allowed value (from the considerations above, and from the perturbative analysis of \cite{Benedetti:2014dra}). We arrive at:
\be \label{eq:cubic-omega}
(4\pi \tau \epsilon^4-V_3\epsilon^2) \omega^3 -2\pi^2\epsilon^4 \omega^2+2\pi^2 \xi = 0\, .
\ee
Although solvable, the general solution to such an equation is not very enlightening due to the presence of several parameters.
We can gain some insight with some further assumptions on the nature of the roots. The discriminant of the cubic equation is:
\be
-108 \pi^4 \xi^2 \epsilon^4 \left(V_3-4 \pi  \tau \epsilon^2\right)^2+64 \pi^8 \xi  \epsilon^{12} \, ,
\ee
and it has the following two roots when viewed as a function of $\xi$:
\be
\xi_1=0\, , \;\;\; \text{and} \;\;\; \xi_2 = \frac{16 \pi^4 \epsilon^8}{27 \left(V_3-4 \pi  \tau \epsilon^2\right)^2} >0 \, .
\ee
For small $\epsilon$ and large $V_3$, the positive root takes a very small value, and therefore we concentrate on the case $\xi>\xi_2$. In such case, the discriminant is negative and therefore the cubic equation has only one real root. Then we can use the representation of the real root in terms of hyperbolic functions, writing:
\be \label{eq:omega-sol}
\omega_0 = 2 \sqrt{\frac{-p}{3}} \cosh \left(\frac{1}{3} {\rm arcosh} \left(\frac{3 q}{2 p}
   \sqrt{-\frac{3}{p}}
   \right)\right)-\frac{a_2}{3 a_3} \, ,
\ee
where:
\be
p = \frac{3 a_3 a_1-a_2^2}{3 a_3^2} = -\frac{4 \pi^4 \epsilon^4}{3 \left(V_3-4 \pi  \tau \epsilon^2\right)^2} \, ,
\ee
\be
q = \frac{27 a_3^2 a_0-9 a_3 a_2 a_1+2 a_2^3}{27 a_3^3} = \frac{16 \pi ^6 \epsilon ^{8}-54 \pi ^2 \xi
 \left(V_3-4 \pi  \tau \epsilon^2\right)^2}{27\epsilon^2 \left(V_3-4 \pi  \tau \epsilon^2\right)^3}\, ,
\ee
and $a_n$ is the coefficient of $\omega^n$ in the cubic equation \eqref{eq:cubic-omega}. Furthermore, in \eqref{eq:omega-sol} we have assumed $p<0$, $q<0$ and $4 p^3 +27 q^2>0$, which are all valid for large enough $V_3$.

Lastly, we take \eqref{eq:hl-graft} with  $A=\sqrt{\xi}/(\epsilon\omega)$ and replace $\omega$ by the solution \eqref{eq:omega-sol}, to obtain a profile which is a function of $\xi$, $V_3$, $\tau$ and $\epsilon$.
Denoting such configuration as $\hat \phi_0(t)$, we are interested in studying:
\be
\mathcal S \equiv \tilde S_{(2+1)-\hl-\textrm{mini}}[\bar\phi_0] - \tilde S_{(2+1)-\hl-\textrm{mini}}[\hat\phi_0]
\ee
as a function of $\xi$, at fixed $\kappa^2$, $V_3$, $\tau$ and $\epsilon$, in order to check if and when the droplet configuration \eqref{eq:hl-graft} dominates  (i.e. $\mathcal S>0$).

More conveniently, we can re-express $\mathcal S$ in terms of the discrete BIB dimensionless parameters, eliminating $\xi$, $V_3$ and $\tau$ by means of the relations:
\be
\frac{16 \pi \xi}{\epsilon^2} = \frac{v b_2}{b_1 m_{min}}\, ,
\ee
\be
\frac{V_3}{(4\pi \epsilon^2)^{3/2}} = \frac{M}{m_{min}^{3/2} v^{1/2}} \, ,
\ee
\be
\frac{\tau}{(4\pi \epsilon^2)^{1/2}} = \frac{T}{ (m_{min} v)^{1/2}} \, .
\ee
One then finds that both $\epsilon$ and $\sqrt{v}$ factor out, so that $\mathcal{S}'\equiv  \kappa^2\mathcal{S}/(\epsilon\sqrt{v})$ depends only on the remaining discrete parameters. In particular, for $m_{min}=1$,  $\mathcal{S}'$ depends only on $M$, $T$ and the ratio $b_2/b_1$.

We thus come to a first conclusion: for fixed $M$ and $T$, the reasoning based on the minimization of the action predicts that if there is a phase transition between a droplet ($\mathcal S>0$) and a correlated fluid phase ($\mathcal S<0$), then the boundary between the two phases is a straight line in the $\{b_1,b_2\}$ plane.
In fact, for fixed $M$ and $T$, $\mathcal{S}'$ is only a function of the ratio $\alpha\equiv b_2/b_1$, and the point at which  $\mathcal{S}'(\alpha_c)=0$ corresponds to the phase transition.
Unfortunately, such an equation cannot be solved in a closed form, and we are limited to a numerical evaluation of $\alpha_c$.
For example, at $M=4000$ and $T=80$ , we find $\alpha_c=1.5$, while  at $M=16000$ and $T=80$ , we find $\alpha_c=7.1$.
We can also check numerically that at large volume $\alpha_c \propto M/T^3$. However, the quantitative predictions about the location of the phase transition should not be taken too seriously, because near the phase transition we expect the fluctuations to become important and affect the transition point.

The above results show that the droplet configuration \eqref{eq:hl-graft} wins over the constant one in a certain range of parameters. What we have not shown is that there are no other solutions that win over both in that same range of parameters. In fact it is plausible that the actual dominant configuration is a smooth version of  \eqref{eq:hl-graft}, i.e. one in which bulk and stalk are joined smoothly. However, we believe that the mechanism described above for our ansatz correctly captures the essential physical features of the problem.
One of the goals of the numerical simulations will be to exclude the existence of very different  configurations winning over the droplet one, and thus to confirm the existence of a droplet phase in the model.

\subsection{From CDT\ to balls-in-boxes}
\label{ssec:hl-to-cdt}

Before we proceed with the numerical investigation, it is worthwhile questioning why one would wish to examine the quantized mini-superspace reduction of a projectable, $z=2$ Ho\v rava-Lifshitz model in the first place.  As anticipated in the introduction, the motivations centre around the potential correspondence between a continuum limit of causal dynamical triangulations (CDT) and Ho\v rava-Lifshitz gravity.

The CDT\ program (in $2+1$ dimensions) proposes a non-perturbative definition for quantum gravity through a path integral of the form:
\begin{equation}
  \label{eq:cdt-pf}
  Z_{\cdt} = \sum_{\Delta\in\mathcal{D}} \frac{1}{|\textrm{aut}(\Delta)|} e^{-S_{\regge}(\Delta)}
\end{equation}
where $\mathcal{D}$ is an ensemble of $2+1$-dimensional piecewise-flat geometries comprised of identical simplicial building blocks (tetrahedra in this case). The members of the ensemble are distinguished by the connectivity among the simplices and, importantly, only configurations satisfying a certain \emph{causal restriction} are permitted \cite{Ambjorn:2000dja}.

The weight contains two factors. $|\textrm{aut}(\Delta)|$ is the size of the automorphism group for the simplicial complex $\Delta$.  For generic configurations, it equals one and so it will be neglected from here on out.  Meanwhile:\footnote{Actually, generically the Regge action for a dynamical triangulation in $(2+1)$ dimensions is:
  \begin{equation}
    \label{eq:cdt-regge-d}
    S_{\regge}(\Delta) = \kappa'_{3}N_{3} - \kappa'_{1}N_{1}\,,
  \end{equation}
  but topological relations allow one to rewrite $N_1$ as a linear combination of $N_3$ and $N_0$.
}
\begin{equation}
  \label{eq:cdt-regge-action}
  S_{\regge}(\Delta) = \kappa_{3}N_3 - \kappa_{0}N_0\,,
\end{equation}
is the Regee action for the geometry in question.  Here, $\kappa_3$ and $\kappa_1$ refer to parameters constructed from the bare Newton's constant and the cosmological constant, while $N_3$ and $N_1$ are respectively the number of tetrahedra and vertices in $\Delta$.

The causality clause in the definition of the ensemble takes several forms.  In its original, global incarnation, it states that permissible simplicial geometries possess a preferred foliation. This is used to ensure a causal structure and, of course, singles out a parameter $t$ that indexes the slices. Recently, a more local expression has been developed and studied \cite{Jordan:2013awa,Jordan:2013iaa,Loll:2015yaa}.  Interestingly, while this new formulation does not rely on a preferred foliation, there are strong indications (stemming from the nascent investigation of the phase structure) that it lies in the same universality class as the original foliated model.

We shall discuss only about the original, global definition for the causal structure. For CDT\ simulations, one generally utilizes a canonical, rather than grand canonical ensemble, meaning fixed $N_3$:
\begin{equation}
  \label{eq:cdt-canonical-pf}
  Z_{\cdt}(T,N_3) = \sum_{\Delta\in\mathcal{D}_{N_3}} e^{\kappa_0N_0}\,.
\end{equation}
There are a number of points worth elaborating on here:
\begin{itemize}
    \item[--] The causal structure within CDT\ allows allows one to differentiate between spacelike and timelike edges.  Thus, the set of $d$-dimensional simplices may be partitioned into subsets indexed by $m\in\{1,\dots,d\}$, indicating that $m$ of its vertices are on one slice, while $d+1-m$ of its vertices are on the later adjacent slice. However, topological relations in $d = 3$  \cite{Ambjorn:2001cv} ensure \eqref{eq:cdt-regge-action} is the most general linear action for such variables.
    \item[--] When performing computer simulations, for reasons of technical simplicity, one imposes a spacetime topology of $S^2\times S^1$; a spherical spatial slice and cyclical time, meaning periodic boundary conditions (in time).  Moreover, as well as fixed 3-volume, one restricts to a fixed number of time steps.   As we commented earlier, this matches a continuum theory wherein the lapse has not been integrated.
    \item[--] The minimal volume of a spatial slice (typically forbidden to be empty by construction) corresponds to the minimal number of triangles needed to triangulate the surface of given topology, while satisfying the appropriate regularity condition (i.e. being a  simplicial manifold). In the case of $S^2$, such number is four (the number of triangles in a tetrahedron).
\end{itemize}

\textbf{Na\"ive continuum limit leads to GR or HL?}
The aim of the game, of course, is to devise a theory some phase of which describes a macrosopic universe governed by Einstein's equations.  Indeed, there is evidence for such a phase within CDT\ and it is being vigorously pursued.

Despite this fact, the discrete theory is written down in terms of certain parameters and with support on certain structures. With regards to the first, this means that of all the ways to take a continuum, it is without doubt true that some stand out more than others purely due to the way the theory is formulated.  This we shall term a na\"ive continuum limit.   With regards to the second point, there is a choice of foliation. This is at odds with the full diffeomorphism invariance of Einstein relativity.  However, there is no reason in a continuum limit (na\"ive or otherwise) that such a preferred foliation could not be washed out and the full space-time diffeomorphism invariance of GR recovered.
Recently, there has been growing evidence that a preferred foliation survives the na\"ive continuum limit of CDT, with the result that the related continuum theories lie in the universality class of HL\ gravity.

There is of course a dearth of analysis with respect to the renormalization group flow of CDT, it has only just begun in earnest \cite{Ambjorn:2014gsa}. So it is wise at this stage to consider any HL\ theory purely in an effective field theory spirit. We have chosen the simplest that possesses certain desirable features. In particular, the \emph{projectability} reduces considerably the number of permissible invariants.  Meanwhile, we have truncated the expansion of our action to order $R^2$, higher order terms being irrelevant at the $z=2$ Lifshitz point, where the theory is renormalizable.

Having said that, non-perturbative calculations in the CDT\ formalism are still time-consuming and even the na\"ive continuum limit is not easy to analyze.  In search of a simpler, more accessible  model that can mimic the essential features, at least for simple observables, one is quite readily led to mini-superspace reductions of these classical gravity theories. These constitute a drastic reduction of degrees of freedom and so, within a CDT-type (that is, lattice) quantization, many of the analytic and computation difficulties are ameliorated. Nevertheless, the spatial volume still acts as an observable that captures non-trivial information from the theory.  With its manifest geometric interpretation, it is interesting to compare its properties within the quantized reduced theory and within CDT\ proper.

Of course, there is no guarantee the quantization of a classically-reduced theory is neatly embedded as some limit of the more general quantized theory.  However, once certain properties have been found in the simpler theory, they serve as motivation to search for them in the more general theory.

\section{Simulations}
\label{sec:sim}

\subsection{Phase diagram}
\label{ssec:ni}

Just as in \cite{Bogacz:2012sa}, we expect a multi-faceted phase structure to emerge and with that in mind, we enlist a number of aptly chosen order parameters to detect phase transitions:
\begin{eqnarray}
  \label{eq:ni-parameters}
  \sigma & = & \left\langle \frac{\sum_{j = 0}^{T-1} m_j^2}{M^2}\right\rangle\,,\\
  \gamma & = & 1 - \left\langle \frac{1}{\min m_j}\right\rangle\,,\\
  \delta & = & 1 - \left\langle \frac{\sum_{j = 0}^{T-1} |m_j - m_{j+1}|}{2M}\right\rangle\,.
\end{eqnarray}
All three parameters lie in the range $[0,1]$, and they each have a simple meaning:
\begin{itemize}
    \item[-] $\sigma$ measures the degree of localization.  If one time slice should possess significantly more volume than the others, then $\sigma$ increases.  Meanwhile, if all times have roughly the same volume, then $\sigma \sim 1/T$.
    \item[-] $\gamma$ signals if any time slice has minimal volume.  $\gamma = 0$ is such an instance (for $m_{min}=1$). Else $\gamma > 0$.
    \item[-] $\delta$ measures the degree of smoothness.  Should the volumes associated to adjacent slices vary greatly, then $\delta \sim 0$. Meanwhile, for smoother samples, $\delta \sim 1$.
\end{itemize}


\begin{figure}[htb]
    \centering
    \includegraphics[scale = 0.5]{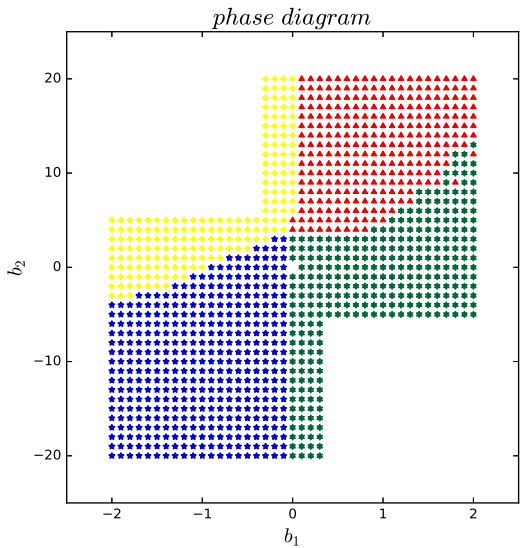}
    \caption{\label{fig:ni-phase} \small{Phase diagram for system with $T = 80$, $M = 4000$: droplet (red triangles), localized (yellow squares), antiferromagnetic (blue pentagons), correlated fluid (green hexagons), and uncorrelated fluid (origin).}}
\end{figure}

\begin{figure}
    \begin{center}
        \begin{minipage}{0.4\textwidth}
            \centering
            \includegraphics[scale = 0.35]{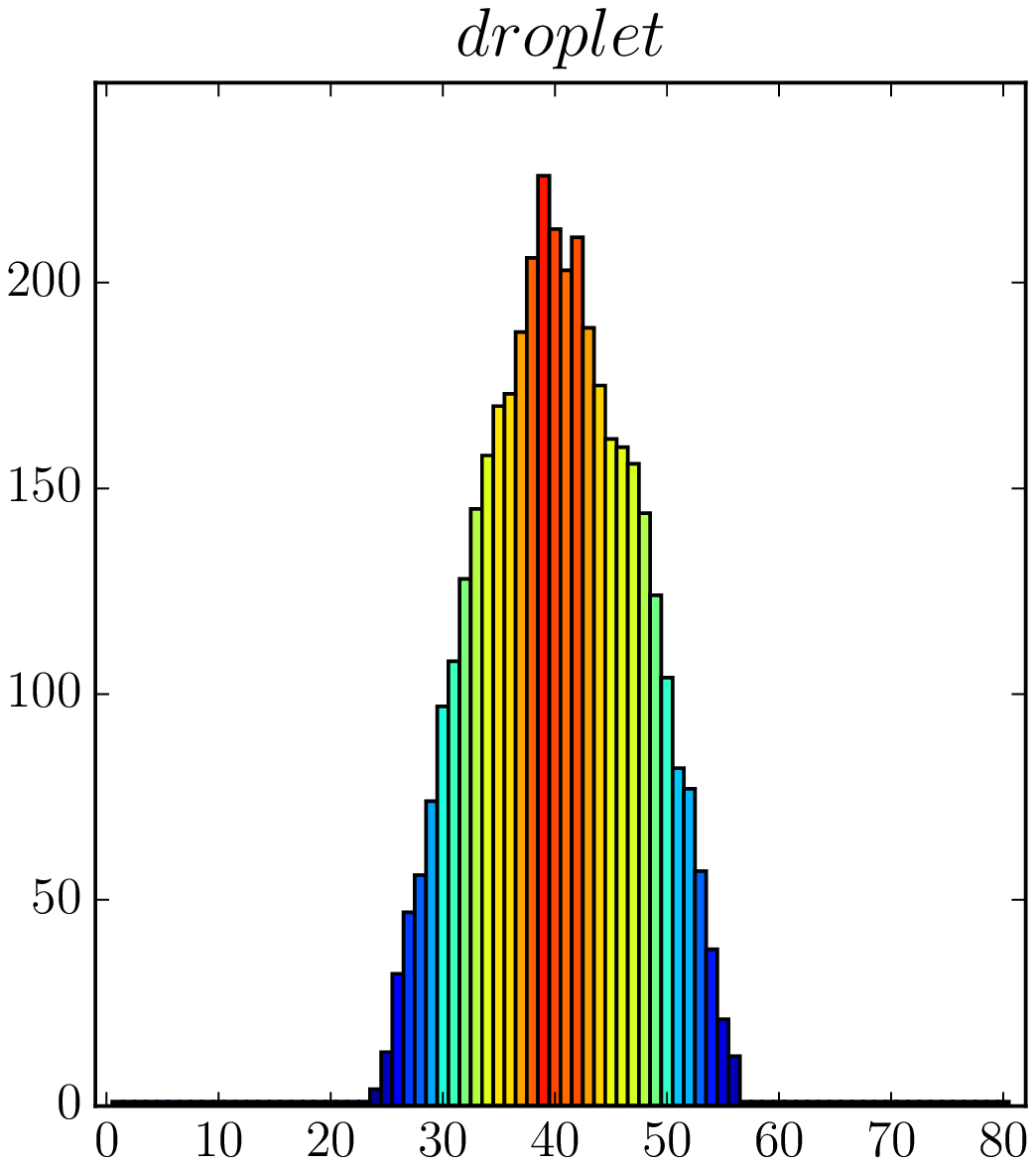}
        \end{minipage}
        \hspace{0.02\textwidth}
        \begin{minipage}{0.4\textwidth}
            \centering
            \includegraphics[scale = 0.35]{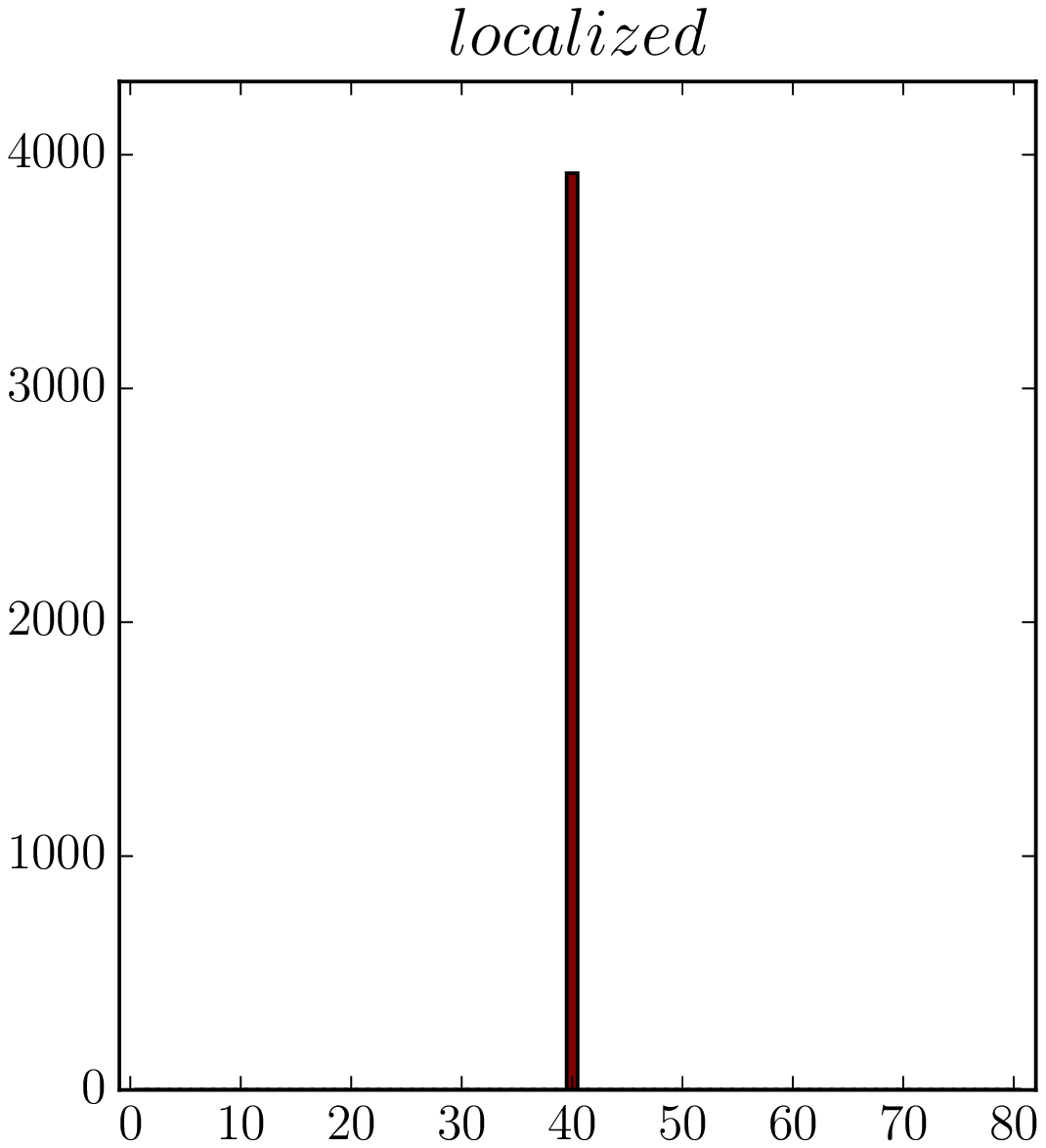}
        \end{minipage}
        \begin{minipage}{0.4\textwidth}
            \centering
            \includegraphics[scale = 0.35]{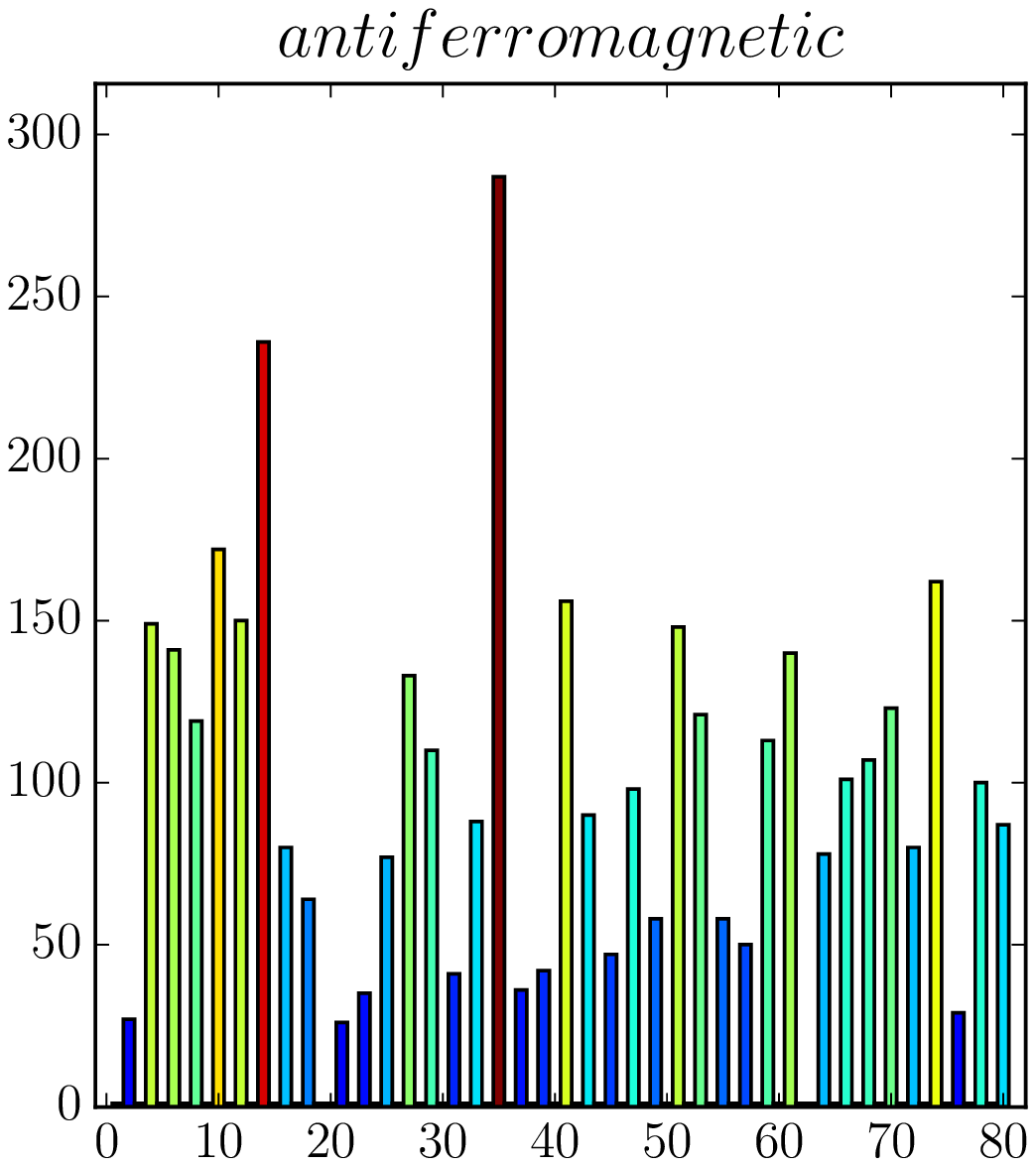}
        \end{minipage}
        \hspace{0.02\textwidth}
        \begin{minipage}{0.4\textwidth}
            \centering
            \includegraphics[scale = 0.35]{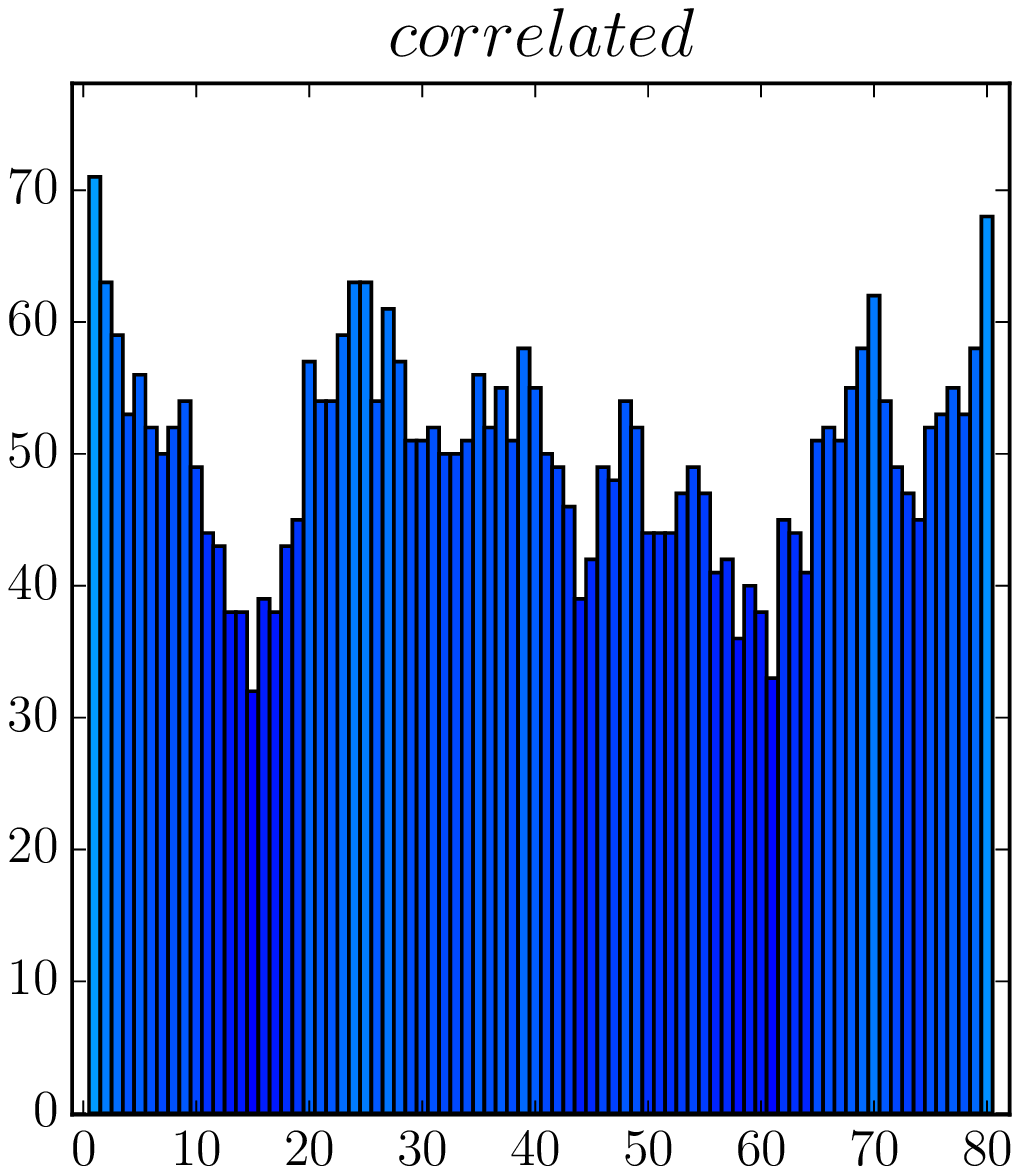}
        \end{minipage}
        \hspace{0.02\textwidth}
        \begin{minipage}{0.4\textwidth}
            \centering
            \includegraphics[scale = 0.35]{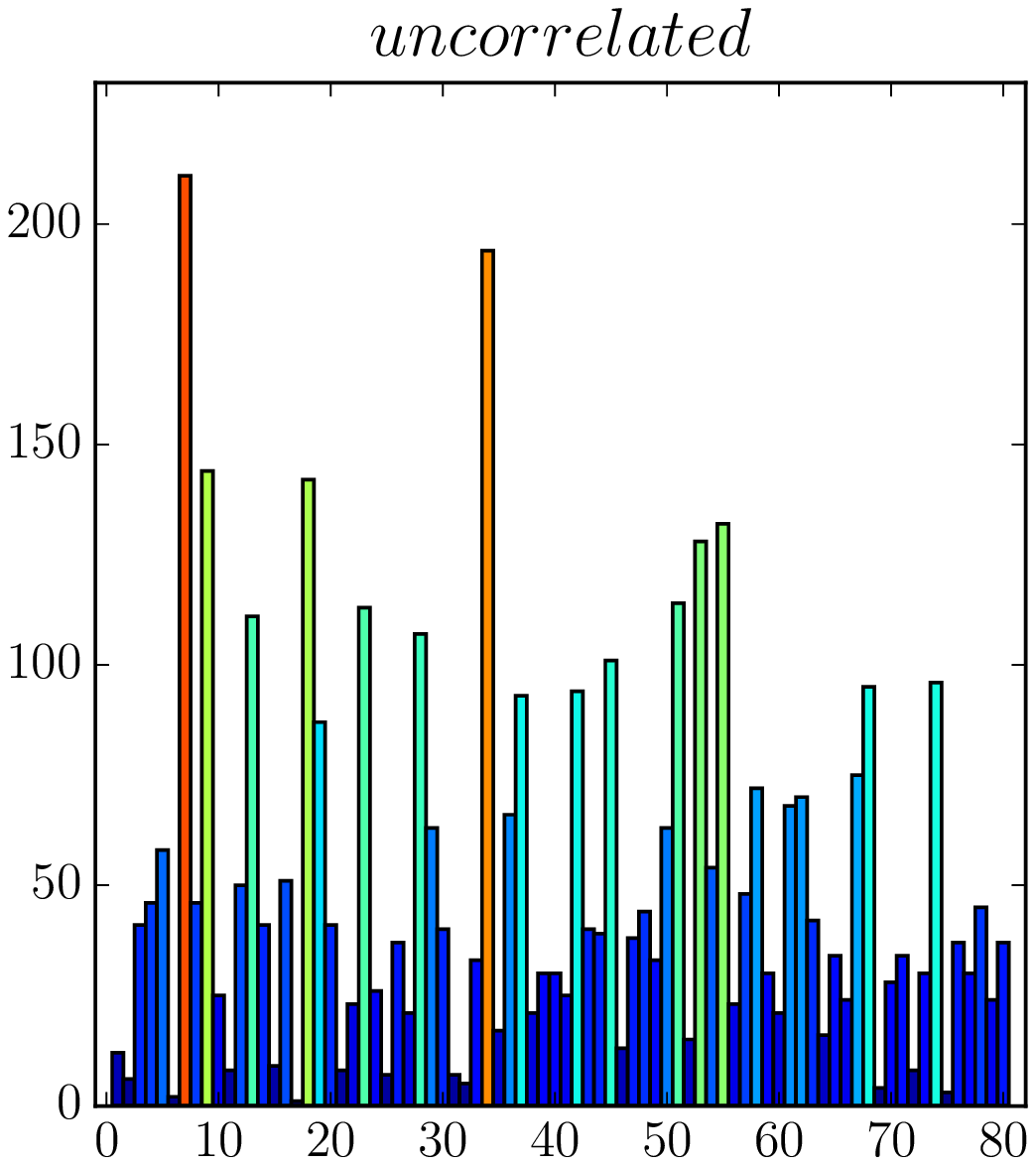}
        \end{minipage}
    \caption{\label{fig:ni-typical} \small{Typical configurations for the various phases (left-to-right): droplet, localized, antiferromagnetic, correlated fluid, uncorrelated fluid.}}
    \end{center}
\end{figure}

We found evidence for five phases, whose location in the phase diagram is illustrated in figure \ref{fig:ni-phase}. Typical configurations for each phase are presented in figure \ref{fig:ni-typical}, and they are characterized as follows:
\begin{description}
    \item[Droplet phase:] $(\sigma \sim 1/W \sim 0, \gamma = 0, \delta \sim 1)$.  This phase occurs for $b_1 > 0$, $b_2 > b_{2,crit}(b_1)$.  For a typical configuration,  the majority of volume condenses onto a contiguous subset of time slices of width $W \gg 1$. The remainder of the volume forms a stalk with each constituent time slice possessing minimal volume.
    \item[Localized phase:] $(\sigma \sim 1, \gamma = 0, \delta \sim 0)$.  This phase exists for $b_1 < 0$, $b_2 > b_{2,crit}(b_1)$. For a typical configuration, all volume resides on one time slice, while all other times possess minimal volume.
    \item[Antiferromagnetic fluid phase:] $(\sigma \sim 2/T \sim 0, \gamma = 0, \delta \sim 0)$. This phase exists for $b_1 < 0$, $b_2 < b_{2,crit}(b_1)$. For finite systems, $\sigma \sim 2/T$. A typical configuration consists of alternating peaks and troughs, where the troughs possess minimal volume.
    \item[Correlated fluid phase:] $(\sigma  \sim 1/T \sim 0, \gamma  \sim 1, \delta \sim 1)$. This phase exists for $b_1 > 0$, $b_2 < b_{2,crit}(b_1)$. For finite systems, $\sigma \sim 1/T$.  A typical configuration consists of the volume approximately evenly allotted to the time slices, with relatively small fluctuations about the mean volume.
        \item[Uncorrelated fluid phase:] $(\sigma   \sim 1/T \sim 0, \gamma  \sim 0, \delta \sim 1/2)$. This phase exists near the origin, i.e. for $b_1\sim b_2 \sim 0$. The values of the exponents at the origin can be computed exactly from the solution at that particular point \cite{Bogacz:2012sa,Waclaw:2007}.
\end{description}


\begin{figure}
    \begin{center}
        \begin{minipage}{0.45\textwidth}
            \centering
            \includegraphics[scale = 0.45]{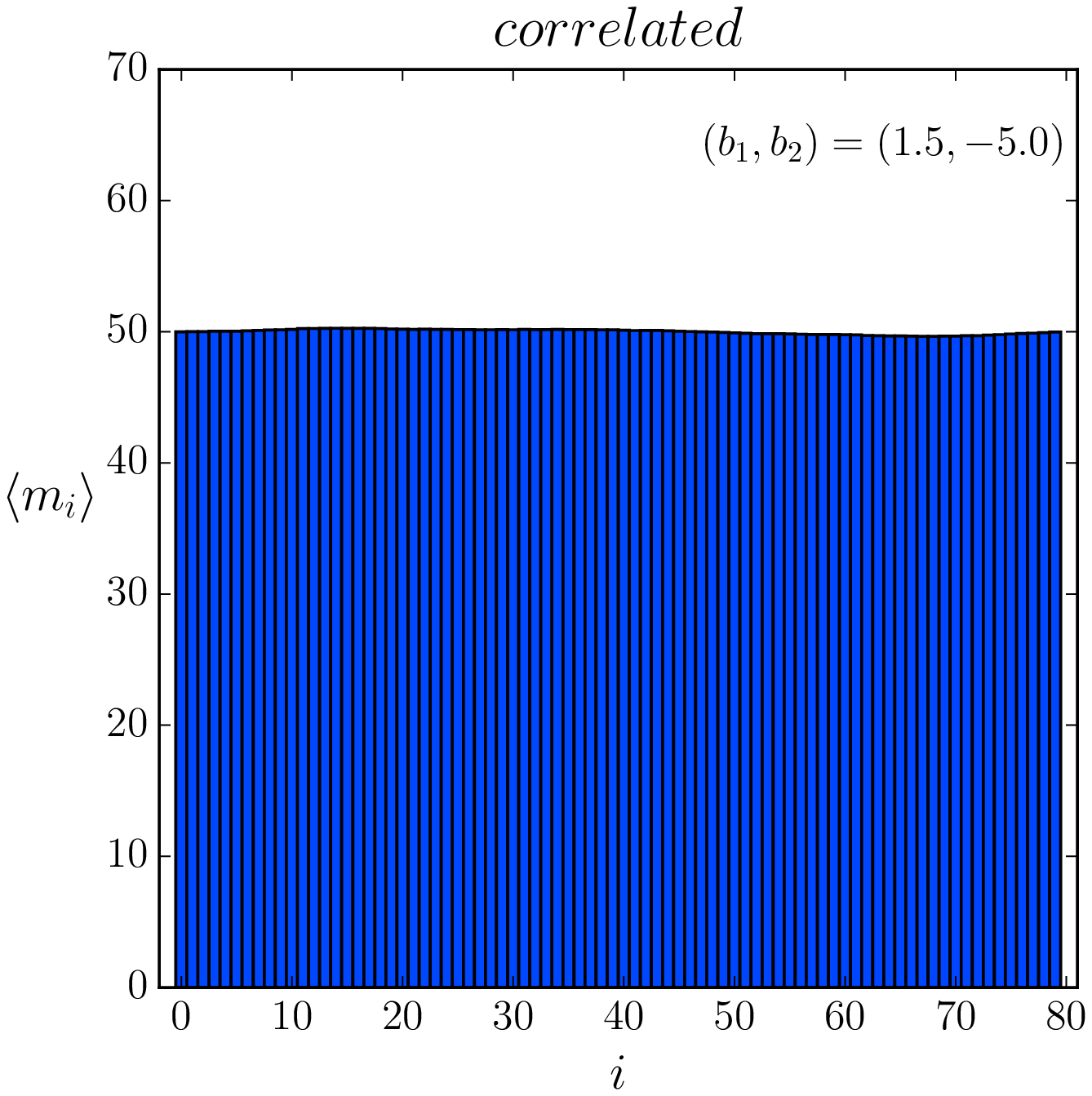}
        \end{minipage}
        \hspace{0.02\textwidth}
        \begin{minipage}{0.45\textwidth}
            \centering
            \includegraphics[scale = 0.45]{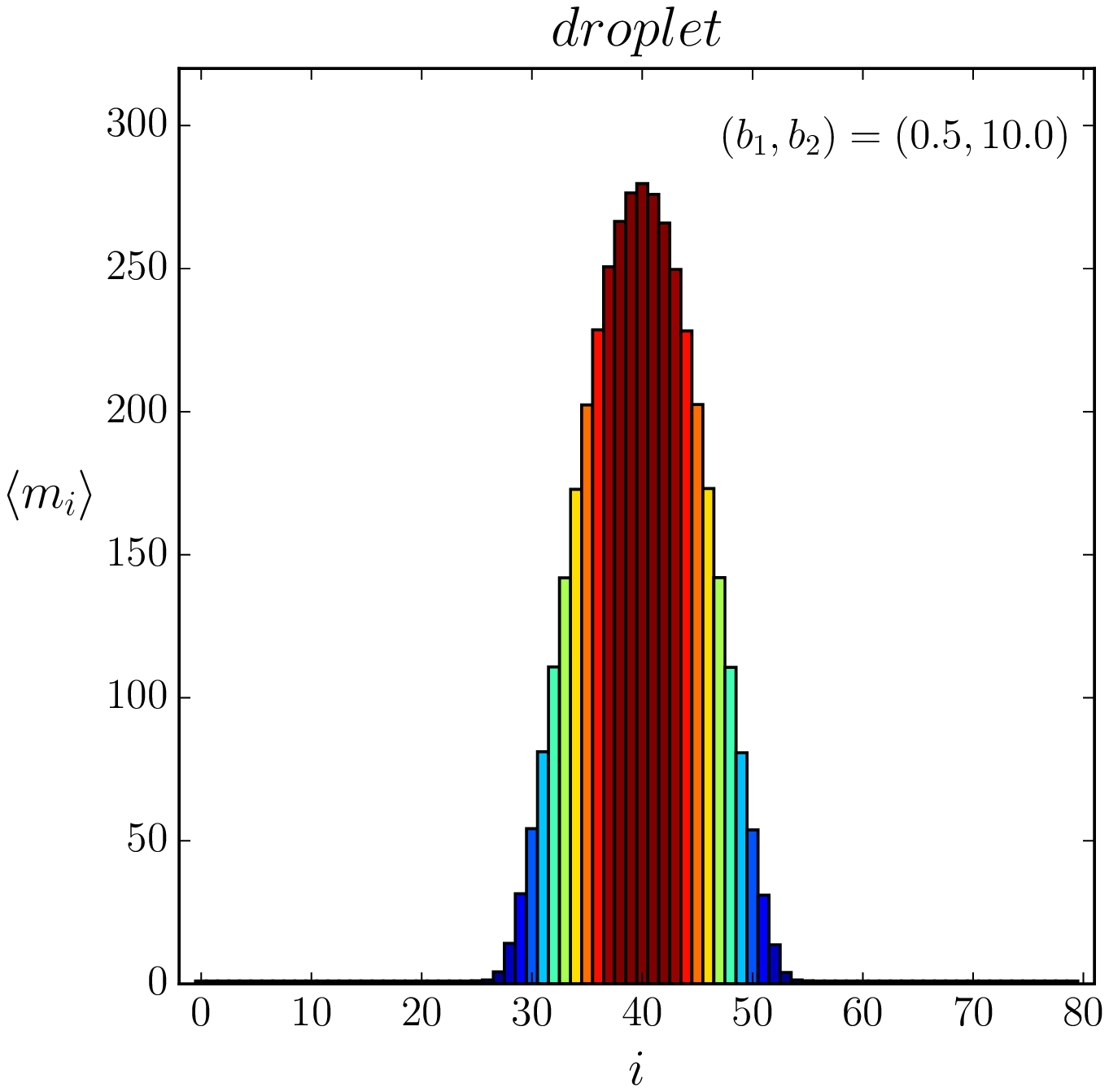}
        \end{minipage}
        \caption{\label{fig:corr-drop_avg} \small{Mean value $\langle m_i\rangle$ as a function of $i$, for non-centred samples in the correlated phase and centred samples in the droplet phase.}}
    \end{center}
\end{figure}


Two of the phases above are in one to one correspondence with the two dominating configurations discussed in  section \ref{ssec:analysis}.
These are the droplet and the correlated phase. Clearly the former should correspond to the droplet configuration \eqref{eq:hl-graft} (see also  figure \ref{fig:corr-drop_avg}), and we will get back to their comparison in section \ref{ssec:bib-vs-cdt}.
That the correlated phase should correspond to the constant configuration is perhaps not completely evident from the plot in figure \ref{fig:ni-typical}, but it certainly should be so from the ensemble average in figure \ref{fig:corr-drop_avg}.
The antiferromagnetic and localized phases appear at negative kinetic term, and for this reason they were not discussed in  section \ref{ssec:analysis}.

\begin{figure}
    \begin{center}
        \begin{minipage}{0.45\textwidth}
            \centering
            \includegraphics[scale = 0.45]{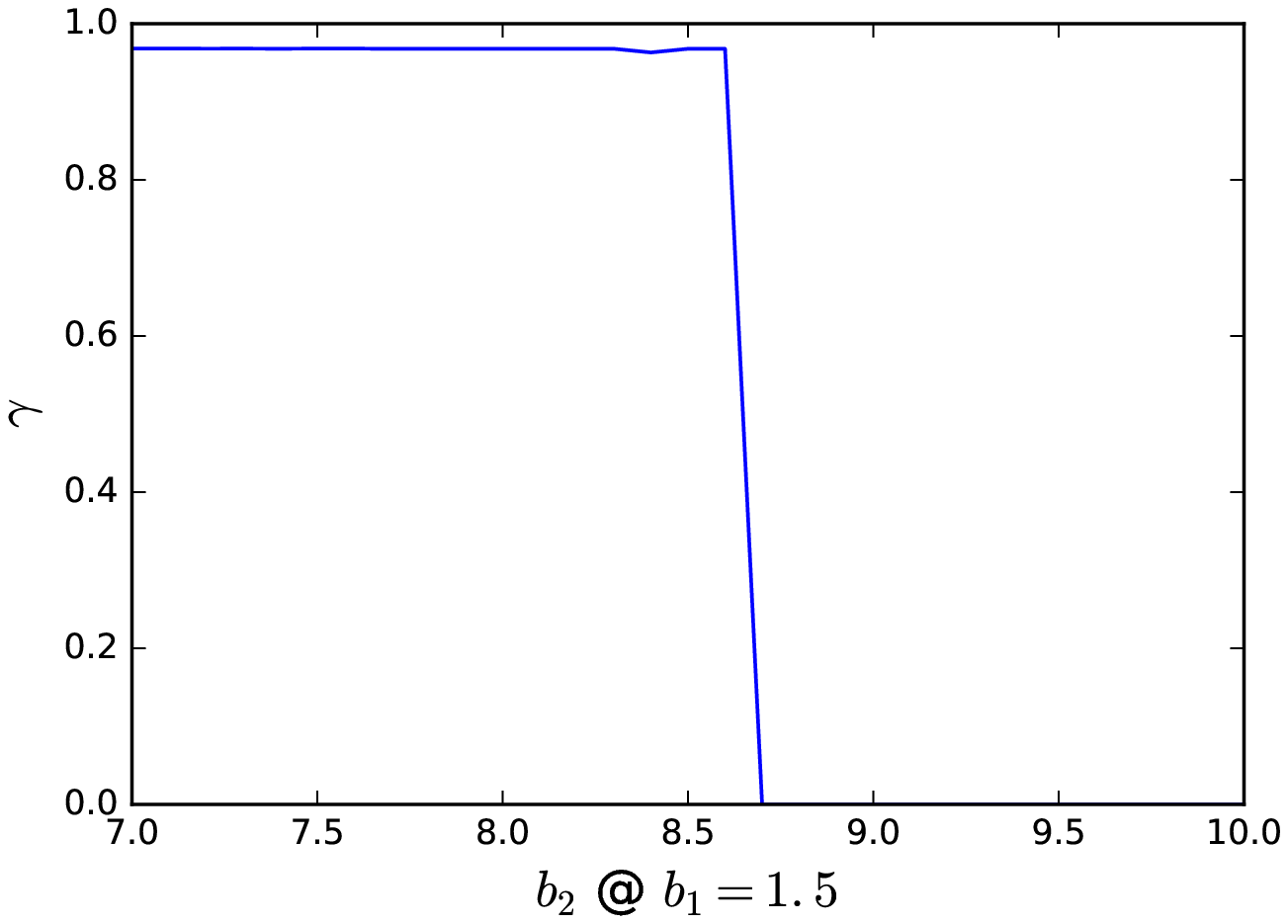}
        \end{minipage}
        \hspace{0.02\textwidth}
        \begin{minipage}{0.45\textwidth}
            \centering
            \includegraphics[scale = 0.45]{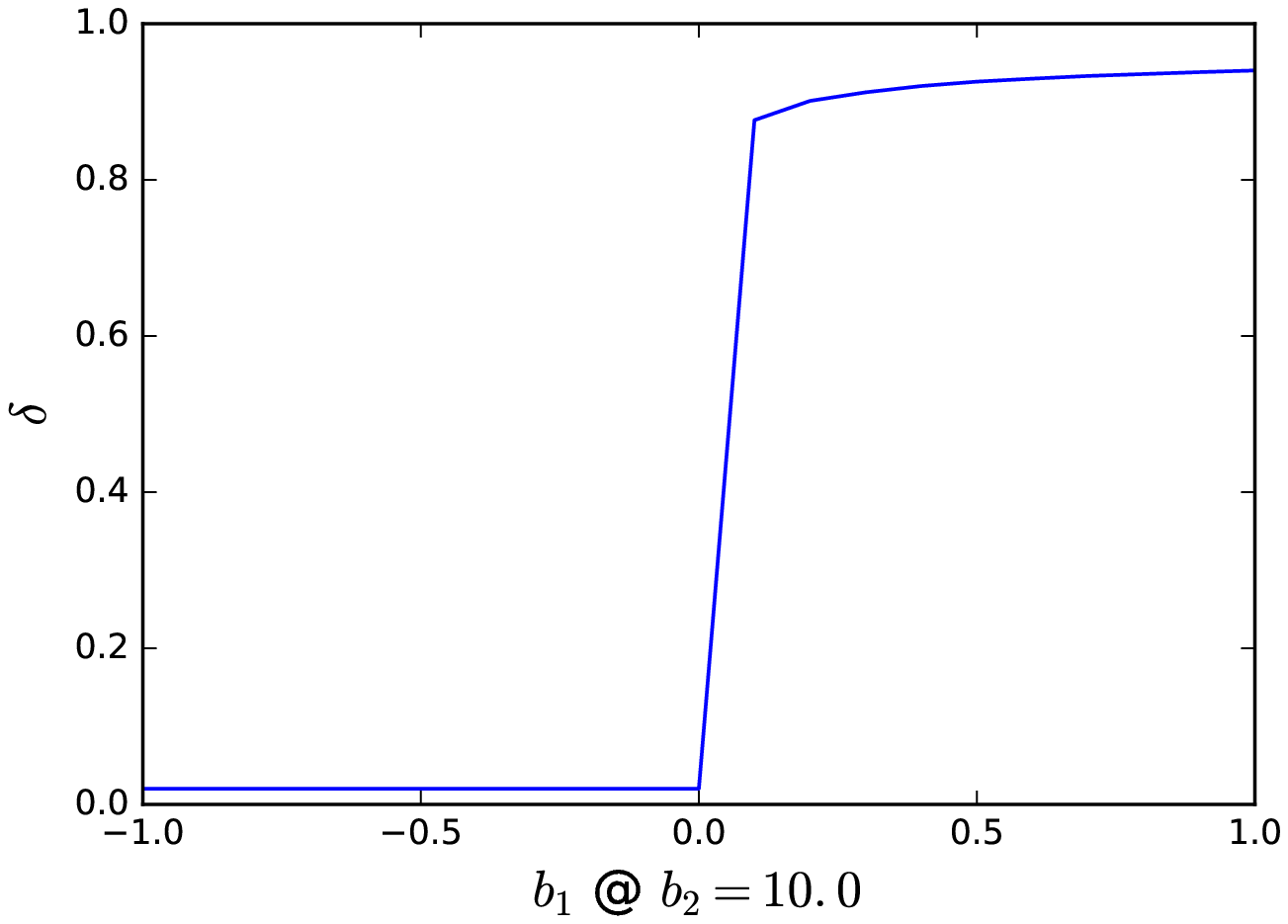}
        \end{minipage}
        \begin{minipage}{0.45\textwidth}
            \centering
            \includegraphics[scale = 0.45]{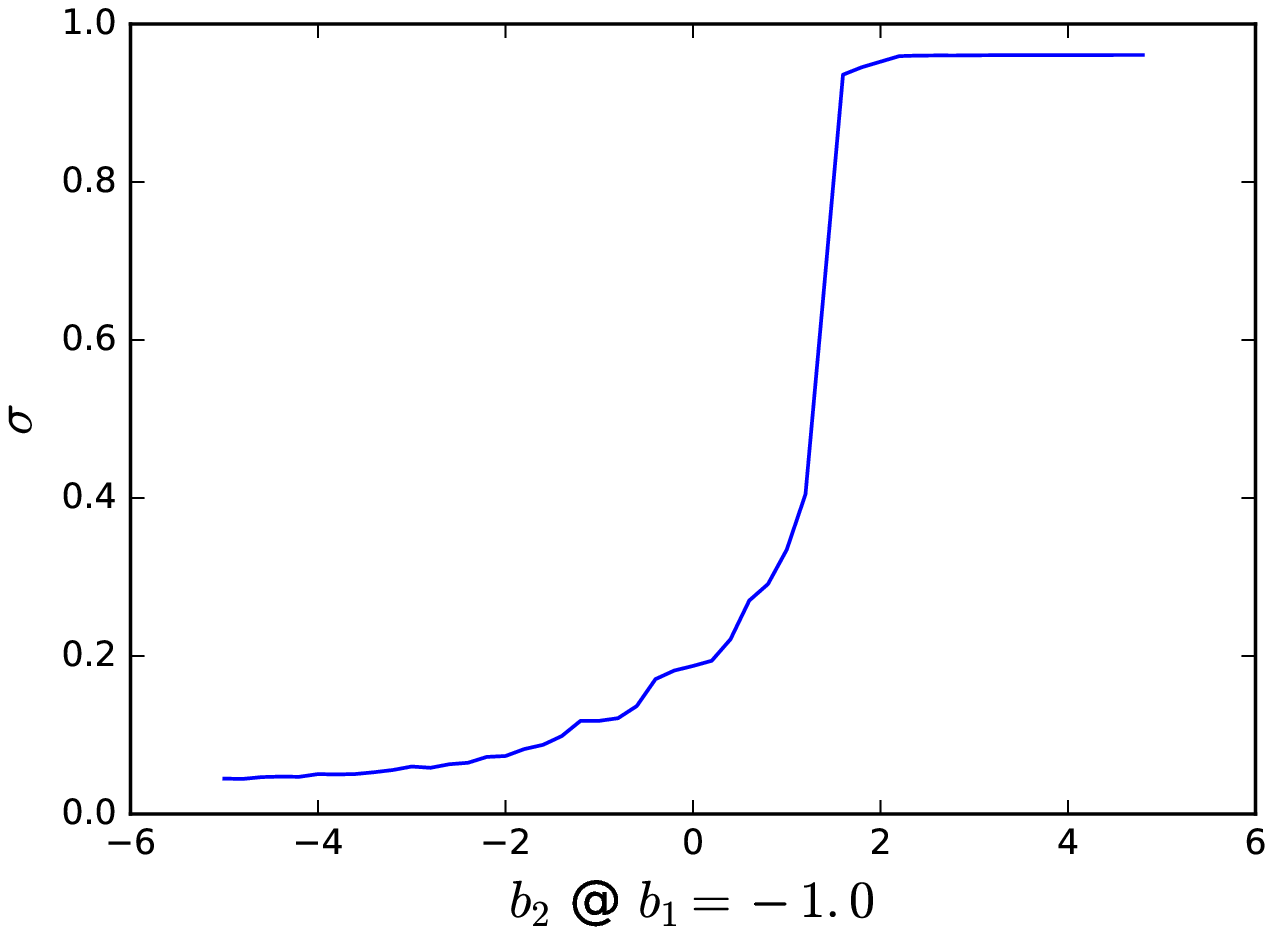}
        \end{minipage}
        \hspace{0.02\textwidth}
        \begin{minipage}{0.45\textwidth}
            \centering
            \includegraphics[scale = 0.45]{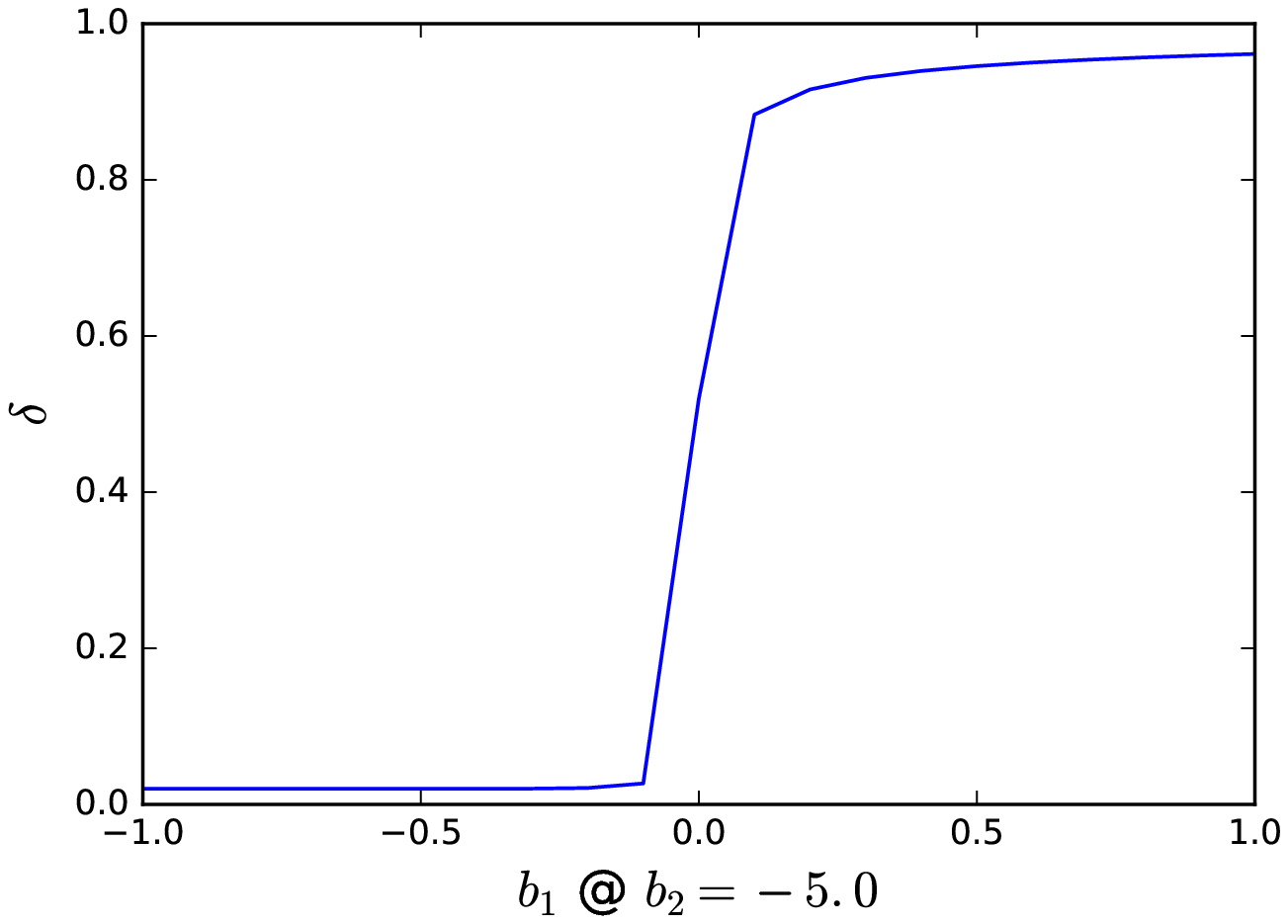}
        \end{minipage}
        \caption{\label{fig:ni-transitions} \small{Plot of the relevant order parameters around the correlated-droplet (top left), localized-droplet (top right), antiferromagnetic-localized (bottom left), and antiferromagnetic-correlated (bottom right)  phase transitions.}}
    \end{center}
\end{figure}

In figure \ref{fig:ni-transitions} we show the behavior of the relevant order parameters around the phase transitions.

\subsection{Mean field sanity check}
\label{ssec:mf}

It is tempting to check that nothing has gone awry by performing a rough mean field approximation to our model along the following lines. One simply evaluates the action in \eqref{eq:bib-pf} on a configuration displaying the rough properties of the dominant configurations in each phase. This provides four functions in the phase parameters $(b_1,b_2)$. One assigns each point $(b_1,b_2)$ to a given phase, dependent on which of the four has least value.

\begin{description}
  \item[Droplet phase:]
    The typical configuration has minimal volume $v_{\textrm{min}}$ at sites outside a droplet of width $W$. For large $M$, the droplet takes on the form $m_j = \frac{M}{W}m\left(\frac{t}{W}\right)$. However, the stalk cannot be neglected due to the potential term in the action.  This leads to:
    \begin{equation}
      \label{eq:mf-droplet}
      \begin{array}{rcl}
        S_{\textrm{droplet}}
        &\sim&
        \dsty
        b_1 \frac{M}{W} \int_0^W \extd t
        \frac
        { 2
          \left[
            m\left(\frac{t+1}{W}\right) - m\left(\frac{t}{W}\right)
          \right]^2
        }
        {
          m\left(\frac{t+1}{W}\right) + m\left(\frac{t}{W}\right)
        }
        - b_2 \frac{W}{M} \int_0^W \extd t
        \frac
        {
          2
        }
        {
          m\left(\frac{t}{W}\right)
          +
          m\left(\frac{t}{W}\right)
        }
        - b_2 \frac{(T - W)}{v_{\textrm{min}}}
        \,,\\[0.5cm]
        &\sim&
        \dsty
        b_1 I_1 \frac{M}{W^2} -
        b_2 I_2\frac{W^2}{M} - b_2\frac{(T - W)}{v_{\textrm{min}}}\,,
      \end{array}
    \end{equation}
    where:
    \begin{equation}
      I_1 = \int_0^1 \extd s \frac{m'(s)^2}{m(s)} \qquad \textrm{and} \qquad I_2 =  \int_0^1 \extd s \frac{1}{m(s)}\,.
    \end{equation}

    Treating the integrals as unknown constants, one minimizes the action with respect to the width $W$ to find a quartic equation in $W$:
    \begin{equation}
      \label{eq:mf-droplet-width}
      W^4 - \frac{M}{2I_2v_\textrm{min}} W^3 + \frac{b_1I_1}{b_2I_2}M^2 = 0
    \end{equation}
    For large $M$, the discriminant is negative indicating two real roots. One finds that, for $b_1 >0$, $b_2>0$, they are positive and behave as:
    \begin{equation}
        \label{eq:mf-real-roots}
        W_0 \sim \left(\frac{b_1I_1m_{min}M}{b_2}\right)^{1/3}, \qquad W_1 \sim \frac{M}{I_2m_{min}}.
    \end{equation}
    The smaller of the two, $W_0$,  has the lesser action, which scales like:\footnote{Meanwhile, the action for the larger root $W_1$ behaves like $S \sim \frac{b_2}{m_{min}^2}M$.
    }
    \begin{equation}
        \label{eq:droplet-action}
        S_{\textrm{droplet}} \sim \left(\frac{b_1b_2^2}{m_{min}^2}\right)^{1/3} M^{1/3}
    \end{equation}

  \item[Localized phase:]
    With essentially all volume ($m_{peak} = M - (T-1)m_{min}$) concentrated at a single site, while the remaining sites possess $m_{min}$, one finds:
    \begin{equation}
      \label{eq:mf-localized}
      S_{\textrm{localized}} \sim 4b_1 M + 4b_1(T+2)m_{min} - b_2 \frac{T-2}{m_{min}},
    \end{equation}
    up to $O(1/M)$ terms.

  \item[Antiferromagnetic phase:]
    A typical configuration has an alternating peak-trough behaviour. On average, these peaks and troughs have volumes $\frac{2M}{T} - m_{min}$ and $m_{min}$, respectively. This leads to a mean field action of the form:
    \begin{equation}
      \label{eq:mf-antiferro}
      S_{\textrm{antiferromagnetic}} \sim 4 b_1 M - 8b_1 Tm_{min} 
    \end{equation}
    up to $O(1/M)$ terms.

  \item[Correlated fluid phase:]
      One assumes that the site volumes vary around a mean volume $\overline{m} = \frac{M}{T}$, with stochastic fluctuations on average of order $\sqrt{\overline{m}}$ (in fact, this is borne out by the numerical simulations). This immediately reduces the action to:
    \begin{equation}
      \label{eq:mf-correlated}
      \begin{array}{rcl}
          S_{\textrm{correlated fluid}} &\sim& \dsty b_1 T - b_1 \frac{T^{3/2}}{\sqrt{M}}, 
      \end{array}
    \end{equation}
    up to $O(1/M)$ terms.

\item[Uncorrelated fluid phase:]
    We do not attempt to incorporate the uncorrelated fluid phase within the mean field analysis.  Although for finite-size systems, one finds it to encompass a small region around the origin in phase diagram, one expects this to shrink to a single point, the origin, in the thermodynamic limit.  At this point, the action vanishes and the system's behaviour is dominated by fluctuations.

\end{description}

The mean field analysis comprises simply of evaluating each of these four actions for different values of the phase space parameters $(b_1,b_2)$. Thereafter, one assigns that point to a given phase, dependent on which of the four has least value.

\begin{figure}[htb]
    \centering
    \includegraphics[scale = 0.5]{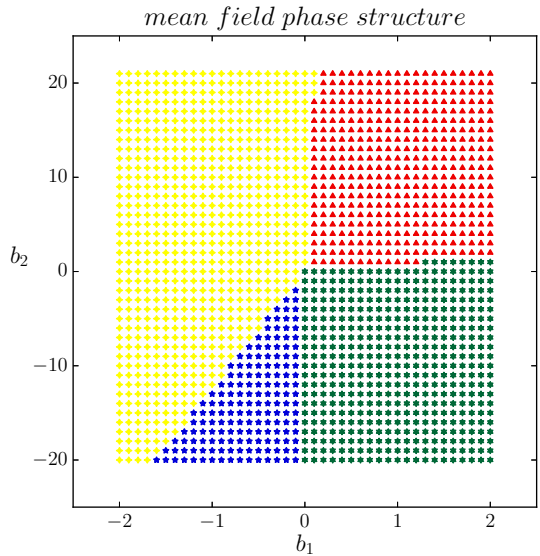}
    \caption{
        \label{fig:mean-field}
    \small{Analytically computed mean field phase diagram for system with $T = 80$, $M = 4000$: droplet (red triangles), localized (yellow squares), antiferromagnetic (blue pentagons), correlated fluid (green hexagons).}}
\end{figure}

As one can see from Figure \ref{fig:mean-field}, one finds qualitatively the same phase diagram as one obtains from the numerical simulations in Section \ref{ssec:ni}. Sanity has been preserved.

Note that at this order, the transition line separating the localized and droplet phases is specified by:
\begin{equation}
    b_{2,c}(b_1) \sim \frac{4m_{min}^2 (3T + 2)}{T-2}b_1 \sim_{T\to \infty} 12 m_{min}^2b_1
\end{equation}
Meanwhile the mean field analysis, for $b_1>0$, also confirms the analysis from section \ref{ssec:analysis}.

\subsection{Comparison to CDT}
\label{ssec:bib-vs-cdt}

As our main motivation was the droplet (also known as extended) phase of CDT, it is now time to actually compare the result of our BIB simulations, with the CDT simulations from \cite{Benedetti:2009ge,Benedetti:2014dra}.

\begin{figure}
    \begin{minipage}{0.45\textwidth}
        \centering
        \includegraphics[scale = 0.32]{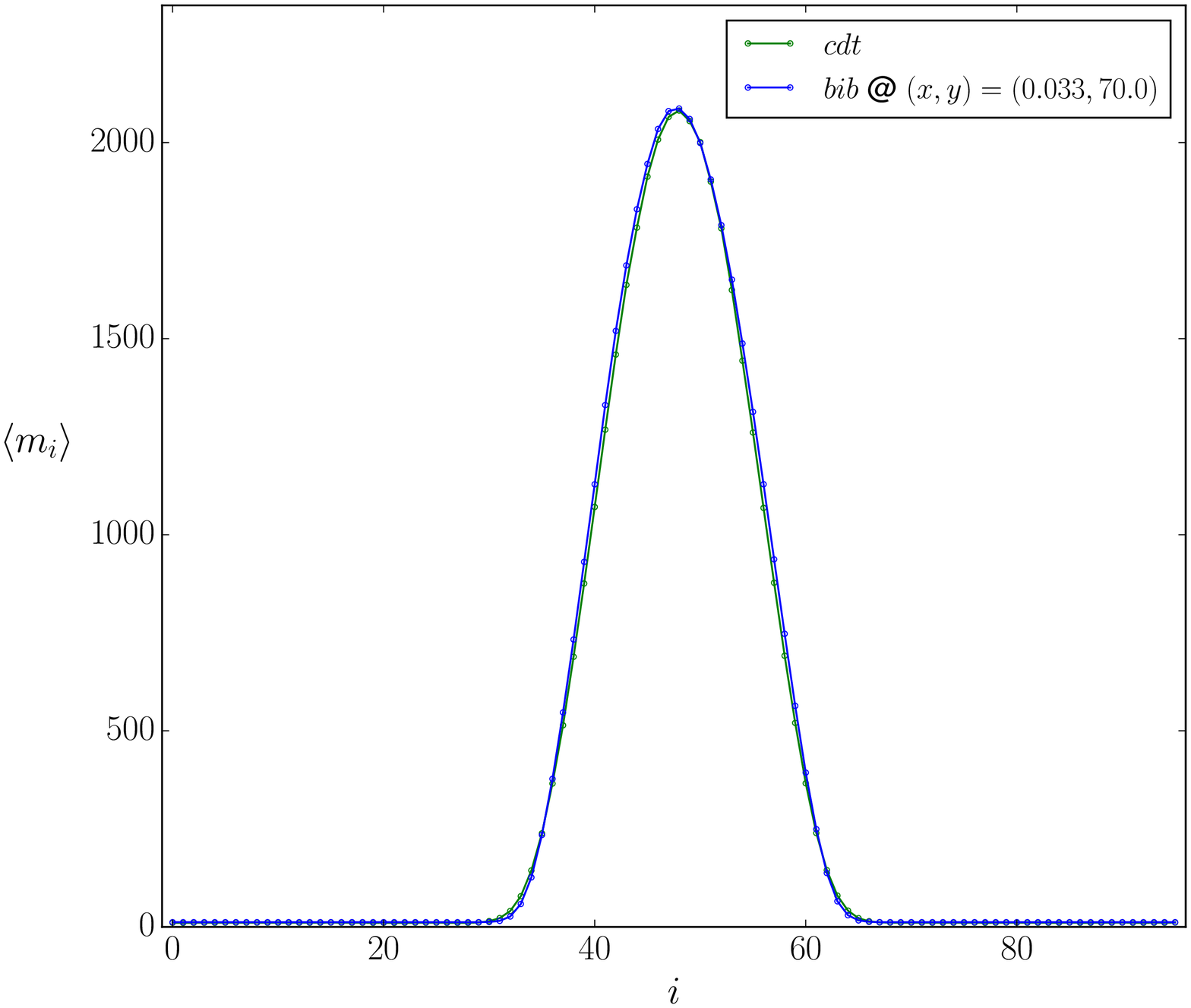}
    \end{minipage}
    \hspace{0.02\textwidth}
    \begin{minipage}{0.45\textwidth}
        \centering
        \includegraphics[scale = 0.32]{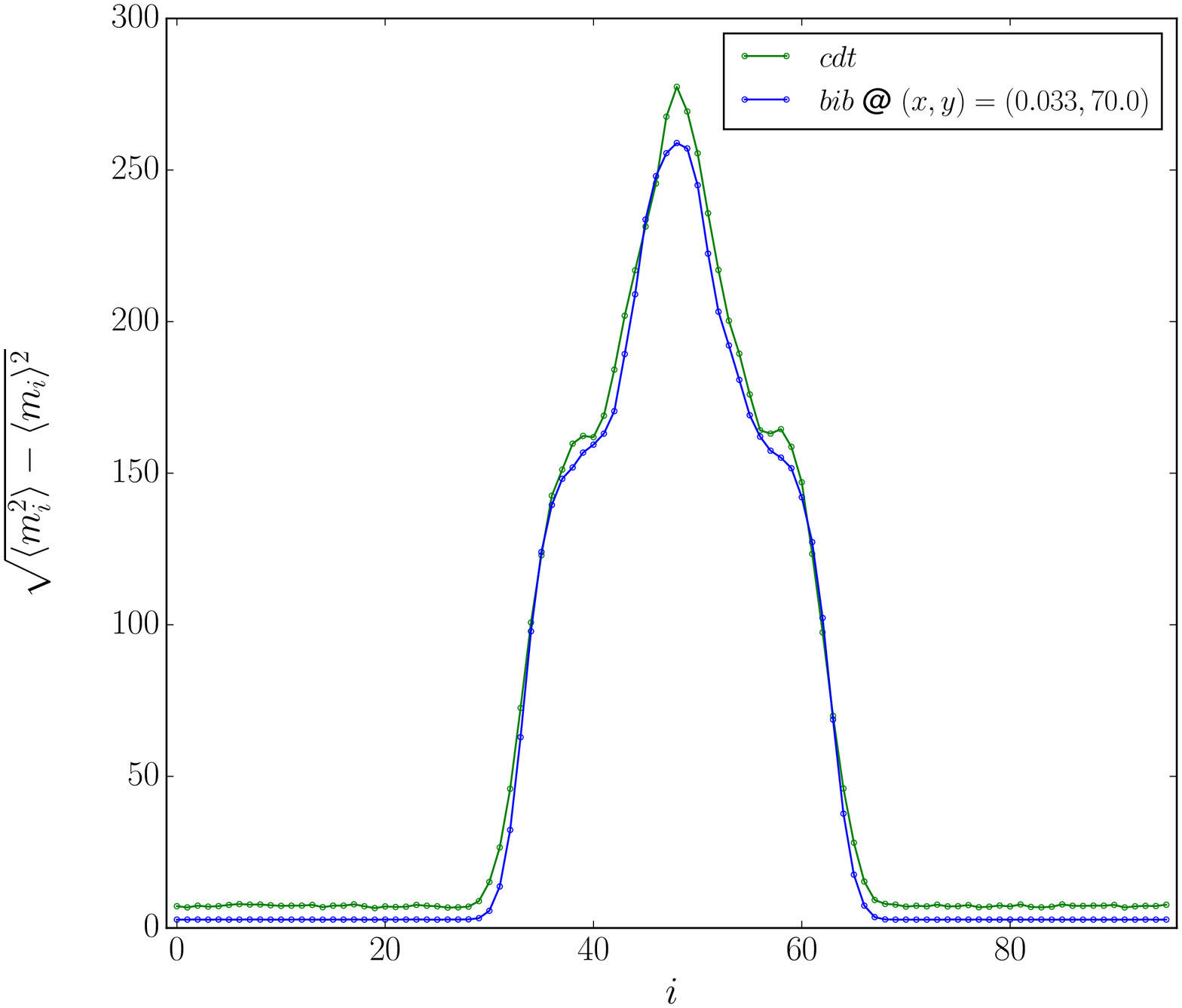}
    \end{minipage}
        \caption{\label{fig:bib-cdt}
        \small{Comparing the BIB droplet and the CDT using the mean volume profile (left) and its standard deviation (right).}}
\end{figure}

We will compare the shape of the respective droplets. To that end, our first task is to shift the center of volume of each sample, within the droplet phase, to some fixed time (the condensate spontaneously breaks time-translation invariance, but as usual we have to pick by hand one ground state out of the continuum of ground states that are related to each other by time translations). We do this following the method described in appendix \ref{app:na}, 
applying it to both the BIB and CDT data. Then, we average over the samples and compare the curves obtained in the two models.
We use as reference data for the CDT side the simulations done with 100k tetrahedra, and total time extension $T=96$.
However, the total volume of the spatial slices in CDT is given by the total number of spatial triangles  $N_2^{(s)}$, and this is roughly one third the number of tetrahedra; more precisely, for our CDT data at  $\kappa_0=5$, we have the expectation value $\langle N_2^{(s)}\rangle =35026$, with standard deviation $\sigma=85$.
Therefore, we repeated the simulations for the BIB model at $M=35026$ and $T=96$ (and $m_{min}=10$, to match the minimum number of triangles observed in a CDT spatial slice), and looked within the droplet phase for the set of parameters $b_1$ and $b_2$ giving rise to a droplet comparable to the CDT one.
The result is shown in figure \ref{fig:bib-cdt}.

Since the shape of the CDT volume profile was already compared to \eqref{eq:hl-graft} in \cite{Benedetti:2014dra}, the excellent agreement between the CDT and BIB data also provides an indirect comparison between the latter and  \eqref{eq:hl-graft}.

We have also compared the fluctuations around the average, i.e. $\sqrt{\langle m_i^2 \rangle - \langle m_i \rangle^2}$, and the result is also shown in  figure \ref{fig:bib-cdt}. The rough shape is again in good agreement, although there are some noticeable differences: the three peaks are flattened down in the BIB case, and the fluctuations in the stalk are more important in the CDT case.
Nevertheless, considering how much simpler is the BIB model in comparison to CDT, it is quite remarkable that even the fluctuations are reproduced to such an extent.

It remains an open question whether the other phases of the BIB model are to be found in CDT.
In (2+1)-dimensional CDT we have in the canonical ensemble only one parameter to tune, and one typically finds only one more phase besides the droplet one \cite{Ambjorn:2000dja}. 
Such a phase is similar to phase ``A'' of (3+1)-dimensional CDT  \cite{Ambjorn:2012jv}, and
as noticed  in \cite{Bogacz:2012sa}, it may correspond to the uncorrelated fluid phase observed in the BIB model.
We will comment in the conclusions about the possible appearance of the correlated fluid phase in extended models of CDT.

\section{Conclusions}
\label{sec:concl}

Our main results are summarized by the figures \ref{fig:ni-phase} and \ref{fig:bib-cdt}. The former shows the rich phase diagram of the BIB model proposed in \cite{Benedetti:2014dra} and obtained here by numerical simulations. In particular, note the presence of a droplet phase, which however would be absent were one to set $b_2=0$, yielding the BIB model related to minisuperspace reduction of general relativity. Figure \ref{fig:bib-cdt} shows the excellent correspondence between the droplet phase of the BIB model and the CDT volume profile. Thus, we have compelling evidence that the effective dynamics of the CDT spatial slices is well captured by the BIB model \eqref{eq:bib-pf}-\eqref{eq:bib-2plus1}, which in turn can be seen as a minusuperspace reduction of the Ho\v{r}ava-Lifshitz model \eqref{eq:hl-action}.
Therefore, our results tighten the link between CDT and HL gravity.

As discussed  in \cite{Benedetti:2014dra}, a further link between CDT and our BIB model, and thus indirectly to HL gravity, comes from the results of \cite{Budd:2013waa}. The authors of that paper studied (2+1)-dimensional CDT  for the case toroidal spatial slices and found no condensation, i.e. no droplet phase. Rather, they only found the equivalent of the correlated phase. This can be understood from our section \ref{sec:model}, because in the toroidal case the minisuperspace reduction of \eqref{eq:hl-action} leads to no potential term. Therefore, one is in the $b_2 = 0$ subspace of the BIB model, and figure \ref{fig:ni-phase} clearly shows how we are in the correlated phase in that case.

An attempt to more directly CDT and HL gravity was made in \cite{Anderson:2011bj}, where the foliated structure of the CDT triangulations was exploited to modify the CDT action in such a way as to correspond to a discretized (2+1)-dimensional HL action.
In particular, the $R^2$ term of \eqref{eq:hl-action} was put in by hand, whereas our results suggest that it arises dynamically in CDT even if we start with the simple Regge action for discrete general relativity. 
And as we argued, it arises with the unstable sign. This means that if in CDT we want to reach an effective BIB model with zero or negative $b_2$ (the stable sign in our conventions), then we need to add to the CDT action an explicit $R^2$ term with a sufficiently large and positive coefficient. In that case, the authors of  \cite{Anderson:2011bj} found a new phase that they called phase ``E''. 
We conjecture that phase E is nothing but the correlated phase and the results presented in  \cite{Anderson:2011bj} appear to corroborate it.\footnote{In particular, note the (almost) absence of minimal slices. The fluctuations around the average of a typical configuration in phase E look larger than what we have in figure \ref{fig:ni-typical} for the correlated phase, but this could be simply due to our choice of phase diagram point for that picture, or to the choice of $\kappa_0$ made in  \cite{Anderson:2011bj}. }

Of course, in order to establish beyond doubt that our BIB model is the effective model of (2+1)-dimensional CDT, it would be important to extract the effective action directly from the CDT data, in a similar fashion to what has been done in \cite{Ambjorn:2011ph,Ambjorn:2012pp} for the (3+1)-dimensional case. We hope to come back to this in the near future.

\appendix
\section{Considerations when performing the simulation}
\label{app:na}

We have performed simulations using Markov-Chain-Monte-Carlo techniques along similar lines to \cite{Bogacz:2012sa}.  To simulate the model at a given point in the phase space $(b_1,b_2)$:

\textbf{Parameter specification:} Apart from the phase space point, one must specify the number of lattice sites $T$, the total volume $M$ and the minimal site volume $m_{min}$.

\textbf{Initialization:} One draws a random initial configuration from the space of BIB model configurations according to a uniform distribution.\footnote{For $M$ units of volume to be distributed among $T$ sites containing at least $m_{min}$ volume units, it is an elementary calculation to show that there are:
\begin{equation}
  \left(
  \begin{array}{c}
    M - (m_{min} - 1)T - 1\\[0.2cm]
   M - m_{min}T
 \end{array}
\right)
\end{equation}
such configurations. Furthermore, it is elementary exercise to devise an algorithm that draws uniformly from this sample space.}

\textbf{Markov Chain:}
One constructs a Markov chain in the space of configurations.
More precisely, say that at some point in the simulation, one is at configuration $A = \{m_0,\dots,m_i,\dots,m_j,\dots,m_{T-1}\}$.
To construct the next potential configuration in the chain, one picks at random two distinct lattice sites, indexed say by $i,j$.
Then, one calculates the relative weight of configuration $B = \{m_0,\dots,m_i - 1,\dots,m_j + 1,\dots,m_{T-1}\}$ with respect to configuration $A$:

\footnotesize
\begin{equation}
  \label{eq:na-relative}
  W(A\rightarrow B) = \left\{
    \begin{array}{lcl}
      \dfrac{g(m_{i-1},m_{i} - 1)
            g(m_{i}-1,m_{i+1})
            g(m_{j-1},m_{j}+1)
            g(m_{j}+1,m_{j+1})
           }
           {
            g(m_{i-1},m_{i})
            g(m_{i},m_{i+1})
            g(m_{j-1},m_{j})
            g(m_{j},m_{j+1})
           }
      &
      \textrm{for}
      &
      j\not\equiv i\pm1 \mod T,\\[0.5cm]

      \dfrac{g(m_{i-1},m_{i} - 1)
            g(m_{i}-1,m_{i+1} + 1)
            g(m_{i+1} + 1,m_{i+2})
           }
           {
            g(m_{i-1},m_{i})
            g(m_{i},m_{i+1})
            g(m_{i+1},m_{i+2})
           }
      &
      \textrm{for}
      &
      j\equiv i+1 \mod T\,,\\[0.5cm]

      \dfrac{g(m_{i-2},m_{i-1} + 1)
            g(m_{i-1} + 1,m_{i} - 1)
            g(m_{i} - 1,m_{i+1})
           }
           {
            g(m_{i-2},m_{i-1})
            g(m_{i-1},m_{i})
            g(m_{i},m_{i+1})
           }
      &
      \textrm{for}
      &
      j\equiv i-1 \mod T\,.

    \end{array}
    \right.
\end{equation}
\normalsize

As one can see, configuration $B$ is generated from $A$ simply by transferring one unit of volume from the site $i$ to site $j$. However, the completion of this transferral is a random variable drawn from a Bernoulli distribution with probability:
\begin{equation}
  \label{eq:na-ap}
  P(A\rightarrow B) = \min\{1,W(A\rightarrow B)\}\,.
\end{equation}
$P(A\rightarrow B)$ is known as the acceptance probability for the move and this method of generating the Markov chain satisfies the necessary requirements of the Metropolis-Hastings algorithm. It is also worth noting that this algorithm automatically conserves the total volume.

\textbf{Restricting the sample space:}
The sample space visited by the algorithm is explicitly restricted to those configurations for which every site satisfies the minimal volume constraint $m_j \ge m_{min}$.  This is effected by immediately rejecting any move that would violate the constraint.

\textbf{Estimating observables/exponents:}
Having generated (and saved) this set of configurations $C$, in principle, it is a simple matter to estimate various physical quantities.  One evaluates the value of the observable for each configuration and performs an arithmetic mean:
\begin{equation}
  \left\langle \mathcal{O}\right\rangle = \frac{1}{|C|}\sum_{c\in C} \mathcal{O}_c
\end{equation}
As with any simulation, there are a number of subtleties that reduce the set of useable configurations to a proper subset of all configurations generated.

\textbf{Subtleties:}
There are an array of technical subtleties that one faces when conducting such a simulation:
\begin{itemize}
    \item[-] To diminish autocorrelation, we thinned the Markov chain by sampling periodically with periods of the order of the total volume.
    \item[-] Large burn-in times are necessary in certain regions of the phase space.
    \item[-] To speed up the algorithm, we precomputed certain pertinent quotients of the transfer matrix:
        \begin{equation}
            \label{eq:na-quotients}
            \begin{array}{rclcrcl}
                \alpha(m_i, m_j) &=& \dfrac{g(m_i, m_j - 1)}{g(m_i, m_j)},&\qquad&\beta(m_i, m_j) &= &\dfrac{g(m_i-1, m_j)}{g(m_i, m_j)},\\[0.5cm]
                \gamma(m_i, m_j) &=&  \dfrac{g(m_i - 1, m_j + 1)}{g(m_i, m_j)},&\qquad& \delta(m_i, m_j) &=&  \dfrac{g(m_i+1, m_j - 1)}{g(m_i, m_j)}.
            \end{array}
        \end{equation}
        for $m_{min} \le m_i, m_j\le m_{max}$, where $m_{max}$ was bounded by available memory. With respect to these auxilliary factors, the acceptance probabilities take the form:
        \begin{equation}
            \label{eq:na-quick}
            W(A\rightarrow B) = \left\{
                \begin{array}{lcl}
                    \dfrac{\alpha(m_{i-1}, m_i) \beta(m_i, m_{i+1})}{\alpha(m_{j-1},m_j+1)\beta(m_j+1, m_{j+1})}
                    &
                    \textrm{for}
                    &
                    j \not\equiv i\pm 1 \mod T,
                    \\[0.5cm]
                    \dfrac{\alpha(m_{i-1}, m_i) \gamma(m_i, m_{i+1})}{\beta(m_{i+1}+1,m_{i+2})}
                    &
                    \textrm{for}
                    &
                    j \equiv i+ 1 \mod T,
                    \\[0.5cm]
                    \dfrac{\beta(m_i, m_{i+1})\delta(m_{i-1}, m_i)}{\alpha(m_{i-2},m_{i-1} + 1)}
                    &
                    \textrm{for}
                    &
                    j \equiv i- 1 \mod T,

                \end{array}
            \right.
        \end{equation}

    \item[-] Within the droplet phase, we centred each of the samples before averaging in order to retain pertinent information about droplet shape.  The periodic boundary conditions suggest embedding the lattice as equidistant points on the origin-centred unit circle in the 2-dimensional plane and calculating the centre-of-mass (that is, centre-of-volume) of the resulting configuration.  The lattice point closest to this 2d centre-of-mass is then designated as central lattice point.
\end{itemize}

\providecommand{\href}[2]{#2}\begingroup\raggedright\endgroup



\providecommand{\href}[2]{#2}\begingroup\raggedright\begin{thebibliography}{10}

\bibitem{Ambjorn:book}
J.~Ambj{\o}rn, B.~Durhuus and T.~Jonsson, {\em {Quantum geometry. A statistical
  field theory approach}}.
\newblock Cambridge Univ. Pr., 1997.

\bibitem{Henson:2006kf}
J.~Henson, {\it {The Causal set approach to quantum gravity}},
  \href{http://arXiv.org/abs/gr-qc/0601121}{{\tt arXiv:gr-qc/0601121}}. in
  `Approaches to Quantum Gravity' ed. D. Oriti, Cambridge Univ. Pr.

\bibitem{Oriti:2011jm}
D.~Oriti, {\it {The microscopic dynamics of quantum space as a group field
  theory}},  in {\em {Proceedings, Foundations of Space and Time: Reflections
  on Quantum Gravity: Cape Town, South Africa}}, pp.~257--320, 2011.
\newblock \href{http://arXiv.org/abs/1110.5606}{{\tt arXiv:1110.5606}}.

\bibitem{Gurau:2016cjo}
R.~Gurau, {\it {Invitation to Random Tensors}},  SIGMA {\bf 12} (2016) 094
  [\href{http://arXiv.org/abs/1609.06439}{{\tt arXiv:1609.06439}}].

\bibitem{Ambjorn:2012jv}
J.~Ambj{\o}rn, A.~Gorlich, J.~Jurkiewicz and R.~Loll, {\it {Nonperturbative
  Quantum Gravity}},  Phys.Rept. {\bf 519} (2012) 127--210
  [\href{http://arXiv.org/abs/1203.3591}{{\tt arXiv:1203.3591}}].

\bibitem{Ambjorn:2004pw}
J.~Ambj{\o}rn, J.~Jurkiewicz and R.~Loll, {\it {Semiclassical universe from
  first principles}},  Phys.Lett. {\bf B607} (2005) 205--213
  [\href{http://arXiv.org/abs/hep-th/0411152}{{\tt arXiv:hep-th/0411152}}].

\bibitem{Horava:2009if}
P.~Horava, {\it {Spectral Dimension of the Universe in Quantum Gravity at a
  Lifshitz Point}},  Phys. Rev. Lett. {\bf 102} (2009) 161301
  [\href{http://arXiv.org/abs/0902.3657}{{\tt arXiv:0902.3657}}].

\bibitem{Ambjorn:2010hu}
J.~Ambj{\o}rn, A.~Gorlich, S.~Jordan, J.~Jurkiewicz and R.~Loll, {\it {CDT
  meets Horava-Lifshitz gravity}},  Phys.Lett. {\bf B690} (2010) 413--419
  [\href{http://arXiv.org/abs/1002.3298}{{\tt arXiv:1002.3298}}].

\bibitem{Benedetti:2009ge}
D.~Benedetti and J.~Henson, {\it {Spectral geometry as a probe of quantum
  spacetime}},  Phys. Rev. {\bf D80} (2009) 124036
  [\href{http://arXiv.org/abs/0911.0401}{{\tt arXiv:0911.0401}}].

\bibitem{Anderson:2011bj}
C.~Anderson, S.~J. Carlip, J.~H. Cooperman, P.~Horava, R.~K. Kommu {\em
  et.~al.}, {\it {Quantizing Horava-Lifshitz Gravity via Causal Dynamical
  Triangulations}},  Phys.Rev. {\bf D85} (2012) 044027
  [\href{http://arXiv.org/abs/1111.6634}{{\tt arXiv:1111.6634}}].

\bibitem{Budd:2011zm}
T.~Budd, {\it {The effective kinetic term in CDT}},  J.Phys.Conf.Ser. {\bf 36}
  (2012) 012038 [\href{http://arXiv.org/abs/1110.5158}{{\tt arXiv:1110.5158}}].

\bibitem{Ambjorn:2013joa}
J.~Ambj{\o}rn, L.~Glaser, Y.~Sato and Y.~Watabiki, {\it {2d CDT is 2d
  Horava-Lifshitz quantum gravity}},  Phys.Lett. {\bf B722} (2013) 172--175
  [\href{http://arXiv.org/abs/1302.6359}{{\tt arXiv:1302.6359}}].

\bibitem{Benedetti:2014dra}
D.~Benedetti and J.~Henson, {\it {Spacetime condensation in (2+1)-dimensional
  CDT from a Ho\v{r}ava-Lifshitz minisuperspace model}},  Class. Quant. Grav.
  {\bf 32} (2015), no.~21 215007 [\href{http://arXiv.org/abs/1410.0845}{{\tt
  arXiv:1410.0845}}].

\bibitem{Horava:2008ih}
P.~Horava, {\it {Membranes at Quantum Criticality}},  JHEP {\bf 03} (2009) 020
  [\href{http://arXiv.org/abs/0812.4287}{{\tt arXiv:0812.4287}}].

\bibitem{Horava:2009uw}
P.~Horava, {\it {Quantum Gravity at a Lifshitz Point}},  Phys. Rev. {\bf D79}
  (2009) 084008 [\href{http://arXiv.org/abs/0901.3775}{{\tt arXiv:0901.3775}}].

\bibitem{Charmousis:2009tc}
C.~Charmousis, G.~Niz, A.~Padilla and P.~M. Saffin, {\it {Strong coupling in
  Horava gravity}},  JHEP {\bf 08} (2009) 070
  [\href{http://arXiv.org/abs/0905.2579}{{\tt arXiv:0905.2579}}].

\bibitem{Blas:2010hb}
D.~Blas, O.~Pujolas and S.~Sibiryakov, {\it {Models of non-relativistic quantum
  gravity: The Good, the bad and the healthy}},  JHEP {\bf 04} (2011) 018
  [\href{http://arXiv.org/abs/1007.3503}{{\tt arXiv:1007.3503}}].

\bibitem{Mukohyama:2010xz}
S.~Mukohyama, {\it {Horava-Lifshitz Cosmology: A Review}},  Class. Quant. Grav.
  {\bf 27} (2010) 223101 [\href{http://arXiv.org/abs/1007.5199}{{\tt
  arXiv:1007.5199}}].

\bibitem{David:2014aha}
F.~David, A.~Kupiainen, R.~Rhodes and V.~Vargas, {\it {Liouville Quantum
  Gravity on the Riemann sphere}},  Commun. Math. Phys. {\bf 342} (2016)
  869--907 [\href{http://arXiv.org/abs/1410.7318}{{\tt arXiv:1410.7318}}].

\bibitem{Ambjorn:1998xu}
J.~Ambj{\o}rn and R.~Loll, {\it {Non-perturbative Lorentzian quantum gravity,
  causality and topology change}},  Nucl. Phys. {\bf B536} (1998) 407--434
  [\href{http://arXiv.org/abs/hep-th/9805108}{{\tt arXiv:hep-th/9805108}}].

\bibitem{Ambjorn:2000dja}
J.~Ambj{\o}rn, J.~Jurkiewicz and R.~Loll, {\it {Non-perturbative 3d Lorentzian
  quantum gravity}},  Phys. Rev. {\bf D64} (2001) 044011
  [\href{http://arXiv.org/abs/hep-th/0011276}{{\tt arXiv:hep-th/0011276}}].

\bibitem{Ambjorn:2002nu}
J.~Ambj{\o}rn, J.~Jurkiewicz and R.~Loll, {\it {3d Lorentzian, dynamically
  triangulated quantum gravity}},  Nucl.Phys.Proc.Suppl. {\bf 106} (2002)
  980--982 [\href{http://arXiv.org/abs/hep-lat/0201013}{{\tt
  arXiv:hep-lat/0201013}}].

\bibitem{Cooperman:2013mma}
J.~H. Cooperman and J.~Miller, {\it {A first look at transition amplitudes in
  (2+1)-dimensional causal dynamical triangulations}},  Class.Quant.Grav. {\bf
  31} (2014) 035012 [\href{http://arXiv.org/abs/1305.2932}{{\tt
  arXiv:1305.2932}}].

\bibitem{Budd:2013waa}
T.~Budd and R.~Loll, {\it {Exploring Torus Universes in Causal Dynamical
  Triangulations}},  Phys.Rev. {\bf D88} (2013), no.~2 024015
  [\href{http://arXiv.org/abs/1305.4702}{{\tt arXiv:1305.4702}}].

\bibitem{Jordan:2013iaa}
S.~Jordan and R.~Loll, {\it {De Sitter Universe from Causal Dynamical
  Triangulations without Preferred Foliation}},  Phys. Rev. {\bf D88} (2013)
  044055 [\href{http://arXiv.org/abs/1307.5469}{{\tt arXiv:1307.5469}}].

\bibitem{Ambjorn:2001br}
J.~Ambj{\o}rn, J.~Jurkiewicz, R.~Loll and G.~Vernizzi, {\it {Lorentzian 3d
  gravity with wormholes via matrix models}},  JHEP {\bf 09} (2001) 022
  [\href{http://arXiv.org/abs/hep-th/0106082}{{\tt arXiv:hep-th/0106082}}].

\bibitem{Benedetti:2007pp}
D.~Benedetti, R.~Loll and F.~Zamponi, {\it {(2+1)-Dimensional Quantum Gravity
  as the Continuum Limit of Causal Dynamical Triangulations}},  Phys. Rev. {\bf
  D76} (2007) 104022 [\href{http://arXiv.org/abs/0704.3214}{{\tt
  arXiv:0704.3214}}].

\bibitem{Benedetti:2013pya}
D.~Benedetti and F.~Guarnieri, {\it {One-loop renormalization in a toy model of
  Horava-Lifshitz gravity}},  JHEP {\bf 1403} (2014) 078
  [\href{http://arXiv.org/abs/1311.6253}{{\tt arXiv:1311.6253}}].

\bibitem{Barvinsky:2015kil}
A.~O. Barvinsky, D.~Blas, M.~Herrero-Valea, S.~M. Sibiryakov and C.~F.
  Steinwachs, {\it {Renormalization of Ho\v{r}ava gravity}},  Phys. Rev. {\bf
  D93} (2016), no.~6 064022 [\href{http://arXiv.org/abs/1512.02250}{{\tt
  arXiv:1512.02250}}].

\bibitem{Bogacz:2012sa}
L.~Bogacz, Z.~Burda and B.~Waclaw, {\it {Quantum widening of CDT universe}},
  Phys.Rev. {\bf D86} (2012) 104015 [\href{http://arXiv.org/abs/1204.1356}{{\tt
  arXiv:1204.1356}}].

\bibitem{Evans-review}
M.~R. Evans and T.~Hanney, {\it Nonequilibrium statistical mechanics of the
  zero-range process and related models},  J. Phys. A: Math. and Gen. {\bf 38}
  (2005) R195.

\bibitem{Jordan:2013awa}
S.~Jordan and R.~Loll, {\it {Causal Dynamical Triangulations without Preferred
  Foliation}},  Phys. Lett. {\bf B724} (2013) 155--159
  [\href{http://arXiv.org/abs/1305.4582}{{\tt arXiv:1305.4582}}].

\bibitem{Loll:2015yaa}
R.~Loll and B.~Ruijl, {\it {Locally Causal Dynamical Triangulations in Two
  Dimensions}},  Phys. Rev. {\bf D92} (2015), no.~8 084002
  [\href{http://arXiv.org/abs/1507.04566}{{\tt arXiv:1507.04566}}].

\bibitem{Ambjorn:2001cv}
J.~Ambj{\o}rn, J.~Jurkiewicz and R.~Loll, {\it {Dynamically triangulating
  Lorentzian quantum gravity}},  Nucl. Phys. {\bf B610} (2001) 347--382
  [\href{http://arXiv.org/abs/hep-th/0105267}{{\tt arXiv:hep-th/0105267}}].

\bibitem{Ambjorn:2014gsa}
J.~Ambj{\o}rn, A.~Gorlich, J.~Jurkiewicz, A.~Kreienbuehl and R.~Loll, {\it
  {Renormalization Group Flow in CDT}},  Class.Quant.Grav. {\bf 31} (2014)
  165003 [\href{http://arXiv.org/abs/1405.4585}{{\tt arXiv:1405.4585}}].

\bibitem{Waclaw:2007}
B.~{Waclaw}, L.~{Bogacz}, Z.~{Burda} and W.~{Janke}, {\it {Condensation in
  zero-range processes on inhomogeneous networks}},  Phys.Rev. {\bf E76}
  (2007), no.~4 046114 [\href{http://arXiv.org/abs/cond-mat/0703243}{{\tt
  arXiv:cond-mat/0703243}}].

\bibitem{Ambjorn:2008wc}
J.~Ambj{\o}rn, A.~Gorlich, J.~Jurkiewicz and R.~Loll, {\it {The Nonperturbative
  Quantum de Sitter Universe}},  Phys. Rev. {\bf D78} (2008) 063544
  [\href{http://arXiv.org/abs/0807.4481}{{\tt arXiv:0807.4481}}].

\bibitem{Gorlich:2011ga}
A.~Gorlich, {\it {Causal Dynamical Triangulations in Four Dimensions}},
  \href{http://arXiv.org/abs/1111.6938}{{\tt arXiv:1111.6938}}.

\bibitem{Ambjorn:2011ph}
J.~Ambj{\o}rn, A.~Gorlich, J.~Jurkiewicz, R.~Loll, J.~Gizbert-Studnicki and
  T.~Trzesniewski, {\it {The Semiclassical Limit of Causal Dynamical
  Triangulations}},  Nucl.Phys. {\bf B849} (2011) 144--165
  [\href{http://arXiv.org/abs/1102.3929}{{\tt arXiv:1102.3929}}].

\bibitem{Ambjorn:2012pp}
J.~Ambj{\o}rn, J.~Gizbert-Studnicki, A.~Gorlich and J.~Jurkiewicz, {\it {The
  Transfer matrix in four-dimensional CDT}},  JHEP {\bf 1209} (2012) 017
  [\href{http://arXiv.org/abs/1205.3791}{{\tt arXiv:1205.3791}}].

\end{thebibliography}\endgroup
\end{document}